# Impact of force field polarization on correlated motions of proteins

*Ana Milinski, Annick Dejaegere, Roland H. Stote\**

Institute of Genetics and Molecular and Cellular Biology, 1 Rue Laurent Fries, 67400 Illkirch-Graffenstaden, France

*Corresponding author: rstote@igbmc.fr

ABSTRACT

Correlated motions of proteins underpin many physiological mechanisms, such as substrate binding, signal transduction, enzymatic activity and allostery. These motions arise from low frequency collective movements of biomolecules and have mostly been studied using molecular dynamics simulations. Here, we present the effects of two different empirical energy force fields used for molecular dynamics simulations on correlated motions – the non-polarizable CHARMM 36m additive force field and the polarizable Drude-2019 force field. The study was conducted on two proteins, ubiquitin - a small protein with a well-described dynamic - and the nuclear receptor protein PPARγ. The ligand binding domain of PPARγ was of particular interest since its function is to regulate transcription through ligand and coregulator protein binding. It has been previously shown that a dynamical network of correlated motions ensures the transmission of information related to PPARγ ligand binding. We present the results of classical MD simulations where we analyze the results in terms of residue fluctuations, residue correlation maps, community network analysis and hydrophobic cluster analysis. We find that RMS fluctuations tend to be greater and correlated motions are less intense with Drude-2019 force field than with the non-polarizable all atom additive force field. Analysis of large hydrophobic clusters in the respective proteins show a greater loss of native contacts in the simulations using the Drude-2019 force field than in the simulations using the all atom force additive force field. Our results provide the first quantification of the impact of using a polarizable force field in computational studies that focus on correlated motions.

**Introduction**

Long-range correlated motions are considered fundamentally important for key functional properties of proteins such as substrate binding, allostery and catalysis[1]. Changes in correlated motions have been associated to the sensing of ligand binding resulting in the propagation of a signal through the protein to transmit information and alter activity. Studies have suggested that correlated motions of secondary structure elements, such as β-sheets, contribute importantly to protein function[2]. For example, PDZ domains are protein interaction modules that recognize short amino acid motifs at the C-termini of target proteins. Ligand binding affects the transfer of binding information to other domains in the context of PDZ-containing multidomain scaffold proteins. In the PDZ domain, correlated motions can lead to the coupling of the N- and C-terminal ends by pathways involving the β-sheets[3]. Correlated motions can be considered as arising from the low-frequency collective movements of residues and it has been suggested that these protein motions are selected by evolution[4,5].

Theoretically, one of the principal methods for studying correlated motions is by molecular dynamics simulations. Molecular dynamics simulations of proteins rely on the use of empirical force fields, which are parameterized using, for the most part, experimental data and quantum mechanical calculations. While this approach has been used with great success over the past decades to study a wide range of topics, there is a constant effort to introduce improvements. One such effort has been to improve the treatment of electrostatic interactions, which in standard classical force fields, are treated by fixed point charges. Efforts by numerous teams have focused on introducing aspects of electronic polarization. One approach characterizes the charge redistribution within each atom, by either induced dipoles[6] or by a Drude oscillator model[7], and the other approach is based on charge flow between atoms, as implemented in the fluctuating charge (FQ) model[8].

The Drude force field[7,9] is a theoretical framework that introduces an auxiliary particle called the "Drude particle", which represents a loosely bound electron that contributes to the atomic polarizability of the atoms. The Drude model is based on the CHARMM all-atom force field [10], although extensive parametrization work has been carried out over the years to improve the balance of energy and forces following the introduction of the Drude particle. A harmonic oscillator function is used to connect the Drude particle to the atom, simulating the restoring force on the electrons. By this approach atomic polarizability is added, allowing for the simulation of electronic response to an external electric field.

The Drude-2019 model has been used and benchmarked for a variety of systems[11] and several reviews are available[12,13]. Though the Drude-2019 model for polarization has undergone extensive development and application, the analysis associated with the applications has largely focused on aspects of structure, energetics and local dynamics. The extent of testing and applications of the Drude-2019 still lags behind that of classical force fields.

In this work, we assess the impact of the polarization on various dynamical properties, with a specific focus on the correlated motions of proteins. We address this question through the study of two proteins, ubiquitin, which contains 76 residues, and the ligand binding domain of the nuclear receptor peroxisome proliferator-activated receptor gamma (PPARγ), which contains 276 residues.

**II. System details**

**Ubiquitin**: Ubiquitin is a small protein that plays a crucial role in various cellular processes, primarily as a regulator of protein degradation. It is found in nearly all eukaryotic cells and is highly conserved across species. Consisting of 76 amino acids, it has a highly conserved three-dimensional structure, with a characteristic beta-grasp fold (Fig. 1), which consists of a 4 stranded anti-parallel sheet and a single helical region with a β(2)-α-β(2) topology[14]. Ubiquitin contains an additional 5th beta strand. The protein has a flexible C-terminal tail, which is involved in the attachment of ubiquitin to target proteins. Ubiquitination is involved in the regulation of various cellular processes, including cell cycle progression, DNA repair, signal transduction, and immune response. Its primary function is to mark proteins for degradation by the proteasome.

Collective motions in ubiquitin have been suggested to play a role in a conformational switch[15]. Furthermore, a correlation network in ubiquitin was identified as spanning the β-strands linking molecular recognition sites[16]. In another study, an allosteric switch governed by a collective motion that affects protein–protein binding was extensively characterized and validated using a combination of techniques, including NMR, X-ray crystallography, computer simulation, and enzyme inhibitor assays. These studies suggested that the loops are involved in a pincer-like movement involving residues in the loop β1-β2 and the loop α1-β3[17,18]. Human ubiquitin has been previously studied by molecular dynamics simulations using the Drude-2019 polarizable force field in the benchmark work of Kognole *et al*[19] and this current study expands upon those results.

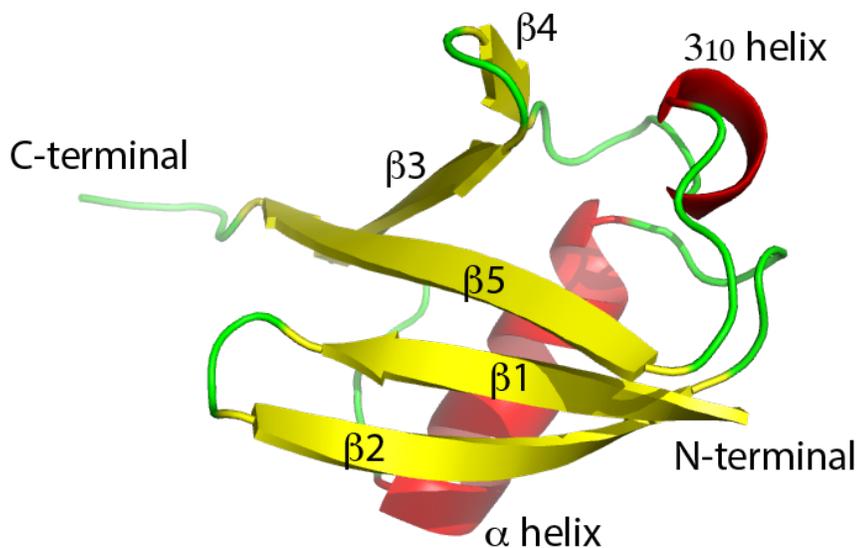

Figure 1: The 3D structure of ubiquitin (1.8Å resolution, PDBID 1UBQ). Shown is the mixed parallel-antiparallel β sheet in yellow, the alpha helices in red, and loop regions in green.

**Peroxisome proliferator-activated receptor gamma**: Peroxisome proliferator-activated receptor gamma (PPARγ) is a ligand-dependent transcription factor belonging to the nuclear receptor superfamily.[20] PPARγ has the common nuclear receptor organization of five conserved domains, including DNA binding (DBD) and ligand binding domain (LBD). Upon signaling events, such as ligand binding, modifications in structure and/or dynamics are relayed to downstream effectors, in particular coregulator proteins. PPARγ is implicated in various diseases, such as obesity, cardiovascular disease and diabetes mellitus, which makes its ligand binding domain (LBD) an important pharmacological target. Various synthetic agonist and antagonist ligands have been shown to regulate PPARγ activity, notably of the family of glitazones[21]. Structures of PPARγ LBD in its apo and corepressor-bound form in complex with a peptide from the NCoR1 corepressor protein are shown in Fig. 2, along with a cartoon drawing labeling secondary structure elements. PPARγ is a larger protein than ubiquitin with a different fold.

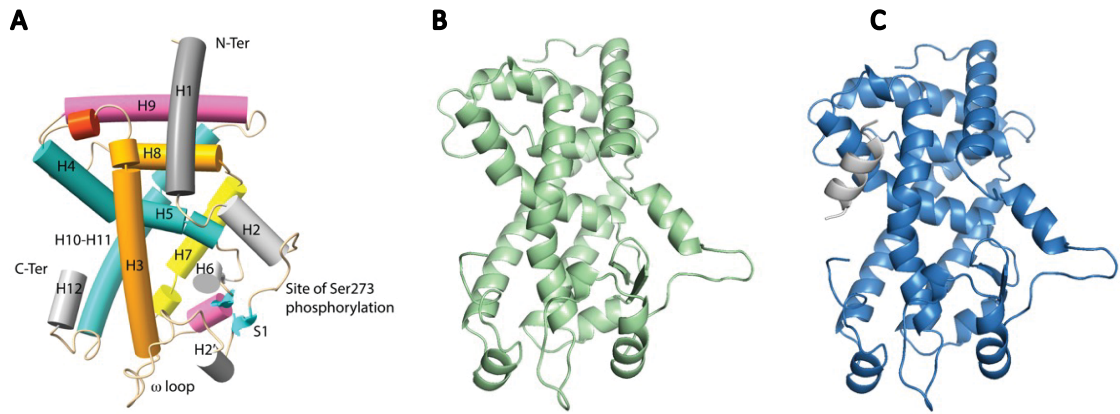

Figure 2. PPARγ ligand binding domain (residues 230 – 505) from the 3.2 Å crystal structure PDBID 7WOX chain B; A) the secondary structure elements are labeled from helix 1 to helix 12, B) the apo form (green) and, C) the same LBD modeled (blue) with the corepressor peptide (gray)

The physiological function of NRs is highly dependent on conformation and structural dynamics modulated by ligand binding. The ligand binding domain acts as a dynamic hub, transmitting binding events to other protein interfaces and domains. PPARγ activation involves a conformational change of the H12 helix at the C-terminal end of the LBD induced by agonist binding[22], where helix H12 undergoes a transition from a flexible ensemble of conformations to a folded conformation stabilized on the core of the LBD. Characterized by numerous crystallographic structures of agonist-bound LBDs, this position of the H12 helix is often referred to as the transcriptionally active conformation[23,24]. In this active conformation, H12, along with helices H3 and H4, constitute a hydrophobic interface called the Activation Function 2 (AF2). This interface serves as a platform for coactivator protein binding and the recruitment of chromatin modulator complexes as well as other components of the basal transcriptional machinery[25]. In contrast to the active conformation of the LBD and H12, the inactive conformation of the receptor is not structurally well described. The crystallographic and computational data suggest an ensemble of conformations for H12, meaning that this region, in the absence of an agonist ligand, is flexible[26]. The important role played by structural dynamics in this protein underpins the need for its further study.

The first study of the functional dynamics of the PPARγ ligand binding domain (LBD) was done by Fidelak *et al*[27]. This study explored the role of allostery in the functioning of the receptor by comparing the LBD in apo and agonist-bound forms. A dynamical pathway linking amino acids that are in topological proximity and at distance was established, explaining correlated motions primarily arising from low-frequency collective motions. The analysis of correlated motions showed coupling between distant regions of the LBD, such as different helices, the N- and C-termini and other physiologically relevant interfaces, such as the co-regulator binding surface and the dimer interface by which PPARγ interacts with its partner, the retinoic X receptor, RXRα. As a consequence, changes in this network could impact the ability of the LBD to bind ligands and coregulators, and by extension the overall function of PPARγ. Correlated motion calculations and network analysis were later done for the full PPARγ/RXRα heterodimer structure in complex with DNA[28]. The results showed the existence of longer range interdomain correlations which were used toward the understanding of allostery in nuclear receptor complexes[28]. Another study showed that phosphorylation of the PPARγ LBD affects the collective motions[29]. Here, we explore the intrinsic dynamics of PPARγ LBD in its apo- and corepressor peptide bound forms using both the classical all-atom additive empirical CHARMM force field [10] and the Drude-2019 force field [9].

We provide quantitative insights into the effects of polarization on the modeling of correlated motions of PPARγ.

**Methods**

Each system was prepared using the PDB Reader and Manipulator option of the CHARMM-GUI web interface[30] to prepare the simulations using the CHARMM all-atom additive force field (AA) using the CHARMM36m all-atom parameter set[10]. The CHARMM-GUI Drude Prepper interface[19] was subsequently used to prepare the systems for simulations using the Drude-2019 (version 2019-H) polarizable force field[9].

For ubiquitin, coordinates were obtained from the 1.8 Å resolution crystal structure (PDB ID 1UBQ)[31]. The protein was solvated in a 64 Å cubic box with 7,921 TIP3P water molecules. The protein carried no net charge and simulations were performed without any neutralizing counterions or added salt. Harmonic constraints were put on the protein and the protein was subjected to an energy minimization followed by a 400 ps equilibration simulation. The constraints were removed and the system was run for 400 ns. The molecular dynamics simulations were done using the NAMD (version 2.14) program under NPT conditions[32]. For the simulations of human ubiquitin using the Drude-2019 force field, the protein was solvated in a cubic box sized $64 \times 64 \times 64$ Å$^3$ with 7,921 "simple water model with four sites and negative Drude-2019 polarizability" (SWM4-NDP) model[33]. The structure was subjected to an energy minimization followed by a 100ps equilibration simulation using a 0.5 fs time step. The shorter than conventional time step was used during this step to better equilibrate the fast degrees of freedom associated with the Drude particles in the Drude-2019 force field. The system was then run under NPT conditions for 200 ns using a 1 fs time step. The particle mesh Ewald method was used to treat the electrostatic interactions.

For the PPARγ ligand-binding domain (residues 230 - 505), we used the 3.2 Å resolution crystal structure of chain B from the PDB file 7WOX[34]. Although one chain in this PDB entry is bound to the antagonist MMT-160, the second chain (chain B) did not show any electron density representing a ligand in the binding pocket, so it was taken to be a structure of the apo protein. The protonation states of the histidine residues of this chain were determined using PROPKA method[35,36] via the poissonboltzmann.org webserver[37] followed by manual verification. The structure was further prepared using the CHARMM GUI interface[30].

The corepressor peptide NCoR ID1 (12 amino acid sequence GLEDIIRKALMG), was added by superposition to a pre-equilibrated structure of 7WOX chain B. The coordinates for the corepressor peptide were taken from the in-house crystallographic structure of a PPARγ mutant complexed to the NCoR peptide resolved by our team. Both the protein/peptide complex and the apo form of the protein were subjected to molecular dynamics simulations using the NAMD program (version 2.14) under NPT conditions[32] and the AA force field. The protocol consists of four steps - first, the protein was fixed, but the water and ions were without constraints. The system was subjected to 1000 steps of steepest descent energy minimization to allow the water and ions to adjust position in response to the presence of the protein. Next, the system was heated up to 600 K, during 23000 steps, again with the protein fixed. This was followed by another energy minimization for 1000 steps and heating to 296.5 K. The constraints on the protein/ligand were removed and the entire system was energy minimized for 2000 steps. Finally, the entire system was heated up to 296.5 K over 15000 steps, followed by an equilibration run of 85 000 steps of dynamics, followed by the production phase; a 1 fs time step was used. The duration of each simulation was 100 stages of $1 \times 10^6$ timesteps, which resulted in a 200 ns - long simulations. The last trajectory frame of this simulation was taken as a starting structure for creating the Drude-2019 and AA models of the PPARγ LBD complexed to the corepressor peptide and the apo form of the LBD by removing the peptide. All simulation starting structures are available from the Zenodo archive site (see Data Availability in Supporting Information).

The AA simulations followed the protocol described above. For the simulations with the Drude-2019 force field, the apo- and corepressor-bound structures were solvated in 100 Å cubic water box using the SWM4-NDP water model. A minimization of 2000 steps was done followed by an equilibration for 200000 steps using the NAMD program with the time step of 0.5 fs. During the production phase, we used a time step of 1fs. The duration of each simulation was 100 ns. Three simulations were carried out for each of the PPARγ LBD systems using both the AA and the Drude-2019 force fields.

For each simulation, the root-mean-square coordinate difference (RMSD) and residue averaged backbone atomic root mean square fluctuations (RMSF) were calculated. The calculated fluctuations were compared to the atomic fluctuations calculated from experimental B-factors. Additional analysis is provided in the Supporting Information.

Cross-correlation coefficients were calculated from the molecular dynamic simulations following the equation:

$$C_{ij} = \frac{\langle \Delta r_i \cdot \Delta r_j \rangle}{\sqrt{\langle \Delta r_i^2 \rangle \langle \Delta r_j^2 \rangle}} \quad (1)$$

where $r_i$ and $r_j$ are the displacements from the mean position of residues $i$ and $j$, respectively. From the $C_{ij}$ correlation coefficients, which are organized as a matrix, a cross-correlation map was calculated using a color-coded 2D representation. In this representation, $C_{ij} = 1$ identifies correlated motions and $C_{ij} = -1$ anti-correlated motions. These values give us information concerning the global collective motions. The $C_{ij}$ correlation coefficients were evaluated over 2 ns segments of the simulations. For each segment, a mean structure was calculated and the $C_{ij}$ correlation coefficients were calculated for the backbone atoms. Correlation maps were obtained by averaging the $C_{ij}$ over all time interval blocks.

Community network analysis was performed using the Bio3D package[38]. Contact maps were produced using an atom-atom distance cut-off of ≤10 Å and the correlated motions were obtained from the molecular dynamics simulations. The Girvan–Newman algorithm[39] as implemented in Bio3D was then used for the community detection. The Girvan–Newman method is a graph-based network approach that is based on the edge-betweenness centrality measure, where the edge-betweenness centrality of an individual residue is defined as the number of the shortest paths connecting other residue pairs that pass through it along the MD trajectory, thus providing an estimate of the influence of this residue on communication, or modularity. Communities of residues are characterized by high modularity values, that is, residues in the same community share dense connections, whereas residues of different communities have sparse or no connections at all. The size of a node is related to the size of a community and a larger sphere depicts a higher number of residues in the node. The edges connect coupled communities, where thicker edges correspond to higher degree of correlation. The correlation threshold for edge detection ($c_{ij}$ *cutoff*) was 0.5. The community map analysis results are depicted using colored spheres mapped on the average 3D structure in tube representation using the VMD software[40].

An analysis of the native hydrophobic contacts and their evolution during the molecular dynamics simulations was done. To initiate the analysis, the Protein Tools website[41,42] was used to define the native hydrophobic clusters of isoleucine (Ile), leucine (Leu) and valine (Val), referred to as (ILV)-clusters, in the initial structures used for the simulations of ubiquitin and PPARγ. Using the CHARMM program, the trajectories were then analyzed to calculate Q, the fraction of native contacts observed during the trajectories of the principal cluster. The van der Waals self-energy of the cluster was also calculated along the trajectory and the results from the different systems were compared. A Wilcoxon rank-sum test was used to test the statistical significance of the results.

**Results and Discussion**

**A. Ubiquitin**

*Drude-2019 FF simulation shows larger RMSD time series values*

For the simulations of ubiquitin, the backbone RMSD time series was calculated over the production phase of the simulations using the initial structure as the reference structure. For each frame in the trajectory, the complexes were reoriented over the entire backbone. Ubiquitin simulated both with the AA force field and with the Drude-2019 force field show a stable RMSD during the trajectories (200 ns for the simulation with Drude-2019 and 400 ns for the simulation without), see Fig. S1. The RMSD values reached average plateau values of 1.8 and 2.8 Å for the AA and Drude-2019 simulations respectively, so the Drude-2019 simulation yields a slightly higher RMSD than the AA simulation. No large-scale conformational changes were observed during either simulation. The 2.8 Å plateau value for the RMSD in the Drude-2019 simulation is very similar to what was observed in the work of Kognole *et al*[19].

*Drude-2019 force field simulations show more significant RMS fluctuations*

The by-residue root mean square fluctuations of the backbone atoms for ubiquitin were obtained from the AA and Drude-2019 force field simulations (Fig. 3). The fluctuations are in good agreement along the entire protein sequence and follow a similar pattern, although the fluctuations from the Drude-2019 simulation are somewhat larger, especially at the level of the helix, the third beta strand and the loop connecting them. The fluctuations from both the AA and the Drude-2019 simulations compare qualitatively well to the fluctuations calculated from the crystallographic B-factors. The profiles obtained compare well to the profiles presented by Kognole *et al*[19].

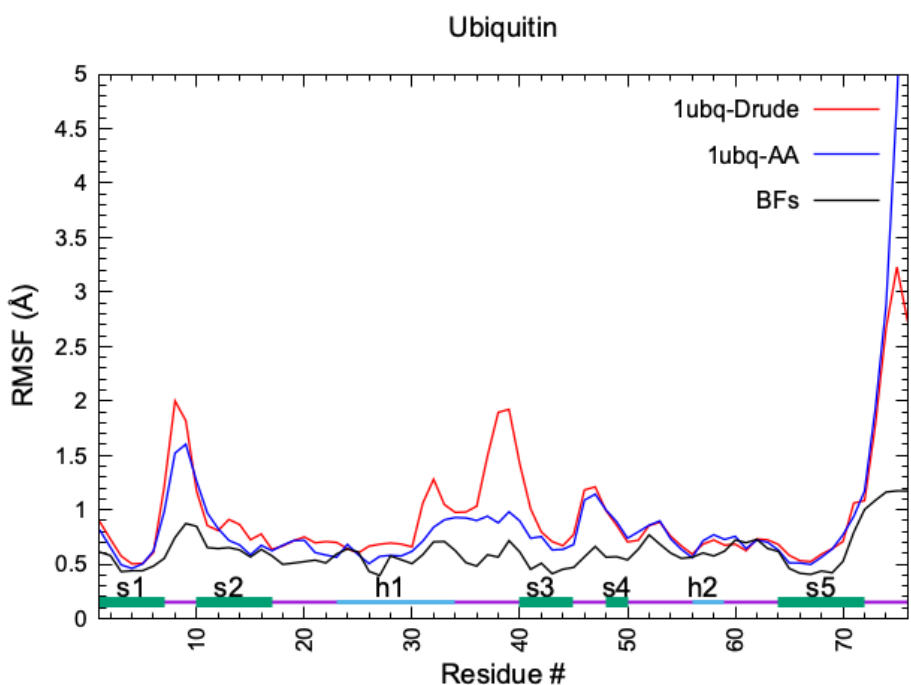

Figure 3: RMSF of backbone atoms averaged by residue for the Drude-2019 simulations (red), the AA simulation (blue) and from the crystallographic B factors (black). The secondary structure elements are indicated.

*Drude-2019 force field simulations show weaker correlated motions*

Correlated motions are important for understanding how the motions in different regions of the protein are coupled to other regions and how they change in response to different perturbations, such as ligand binding. Changes in the correlated motions can effectively occur over long distances. Changes in fluctuations and correlations can be linked to the propagation of allosteric signals through changes in entropy, even in the absence of conformational changes[43]. It is therefore important to identify the residues involved in this transmission of structural dynamic information. This information can be obtained by calculating the cross-correlations, which complement the fluctuation analysis presented above by providing information on correlated motions as calculated by Eq. 1. From the $C_{ij}$ correlation coefficients, which are organized as a matrix, a cross-correlation map is calculated using a color-coded 2D representation. These calculations find use in many different applications[3,27,44].

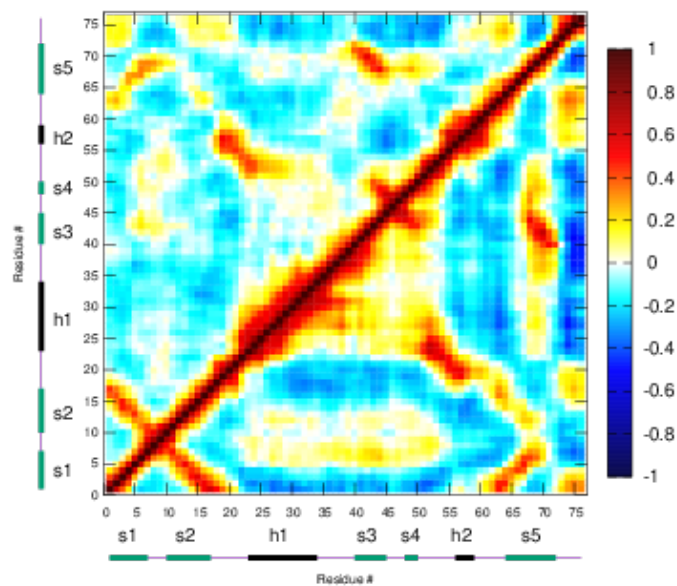

Figure 4: Correlated motions of ubiquitin calculated from the molecular dynamics simulations from the Drude-2019 simulation (upper triangle), compared to the AA simulations (lower triangle). Correlated motion maps are represented with a color code related to the sign and intensity of correlations (ranging from dark blue for perfect anticorrelations to dark red for perfect correlations). The secondary structure elements are indicated.

The correlated motions were calculated from the AA and Drude-2019 molecular dynamics simulations and presented in Fig. 4; their values range from −1 for perfectly anticorrelated motion to +1 for perfectly correlated motion, which is found for the self-correlations of each atom. The upper triangle corresponds to the correlated motions for ubiquitin calculated from the Drude-2019 simulation, while the lower triangle corresponds to the correlation map for ubiquitin calculated from the simulation using the AA force field. We see in the lower triangle the presence of strong correlated motions involving the β strands and α helices. Particularly noteworthy are the motions involving the β-sheet. Correlations between strands β1, β2, β5, between β5 and β3, and between β3 and β4 are observed in both the AA and Drude-2019 simulations, but they are markedly less intense in the Drude-2019 simulations. These correlations correspond to those analyzed by NMR[16], where it was shown that correlated motions extend over 15 Å across the ubiquitin β-sheet. The introduction of polarization via the Drude-2019 model thus does not disrupt the general correlation pattern, but it affects the intensities of the correlated motions.

A scatterplot of the calculated coefficients from the classical and Drude-2019 simulations is shown in Fig. 5, where an overall linear correlation between the two sets of data exists. The results also show that the values from the Drude-2019 simulations are lower (in absolute value) than the value from the AA-simulations (Fig. 5). Out of the 2850 cross-correlation coefficients computed for ubiquitin, 382 differ by more than 0.2 in absolute value between the two data sets. A visual inspection of the data indicates that particularly large discrepancies (0.3 and above) correspond to correlations involving amino acids 37 to 42, which are in the loop between the helix and beta-strand 3. In this region, the rms fluctuations also differ significantly between AA and the Drude-2019 data. A few large differences also involve amino acids in other loops or solvent exposed regions. In Fig. 5B, the probability distribution of the correlation coefficients from the AA simulations is broader and extend to more extreme values than in the Drude simulations.

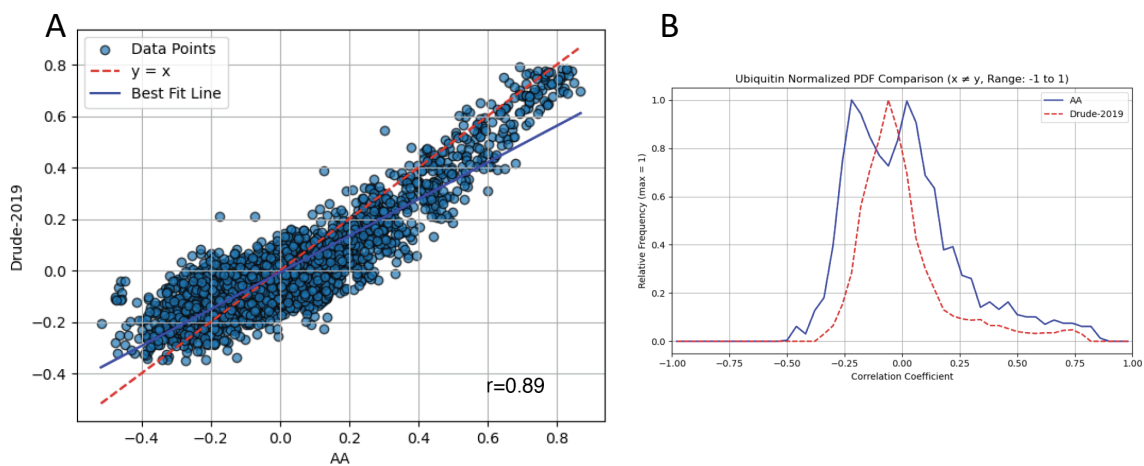

Figure. 5 (A) Scatterplot of the correlation coefficients used to calculate the correlation maps from the molecular dynamics simulation of ubiquitin; from the AA simulations (x-axis) vs from the Drude-2019 simulations (y axis). The dashed line shown in red is x=y; the solid blue line is the best fit to the data excluding the self-correlation coefficients which equal 1, (B) the probability distribution of the correlation coefficients from the AA simulation (solid blue) and the Drude-2019 simulation (dashed red)

To further interpret the consequences of the Drude-2019 force field on long-range correlated motions, we performed a community network analysis (CNA). Maps from a CNA are derived from a functional clustering of correlated motions obtained from MD simulations. It has been shown that this type of analysis can be used to interpret long range communication and dynamic allostery of proteins[45,46]. Community maps can help interpret how different parts of proteins move together and how changes in one part of a protein can affect the dynamics of distant sites. Communities highlight regions of the protein that exhibit collective movements and may represent functionally important domains or allosteric communication pathways. We obtained coarse grained networks of dynamically coupled communities using the Bio3D package[38] and the correlation matrices calculated from our MD simulations. The results of our analysis are mapped onto the average 3D protein structure in tube representation. Communities are depicted as colored spheres (nodes), where the radius of a node is proportional to the size of its community. Lines (edges) connect coupled communities, where the thickness of an edge is proportional to the degree of correlation between the two nodes.

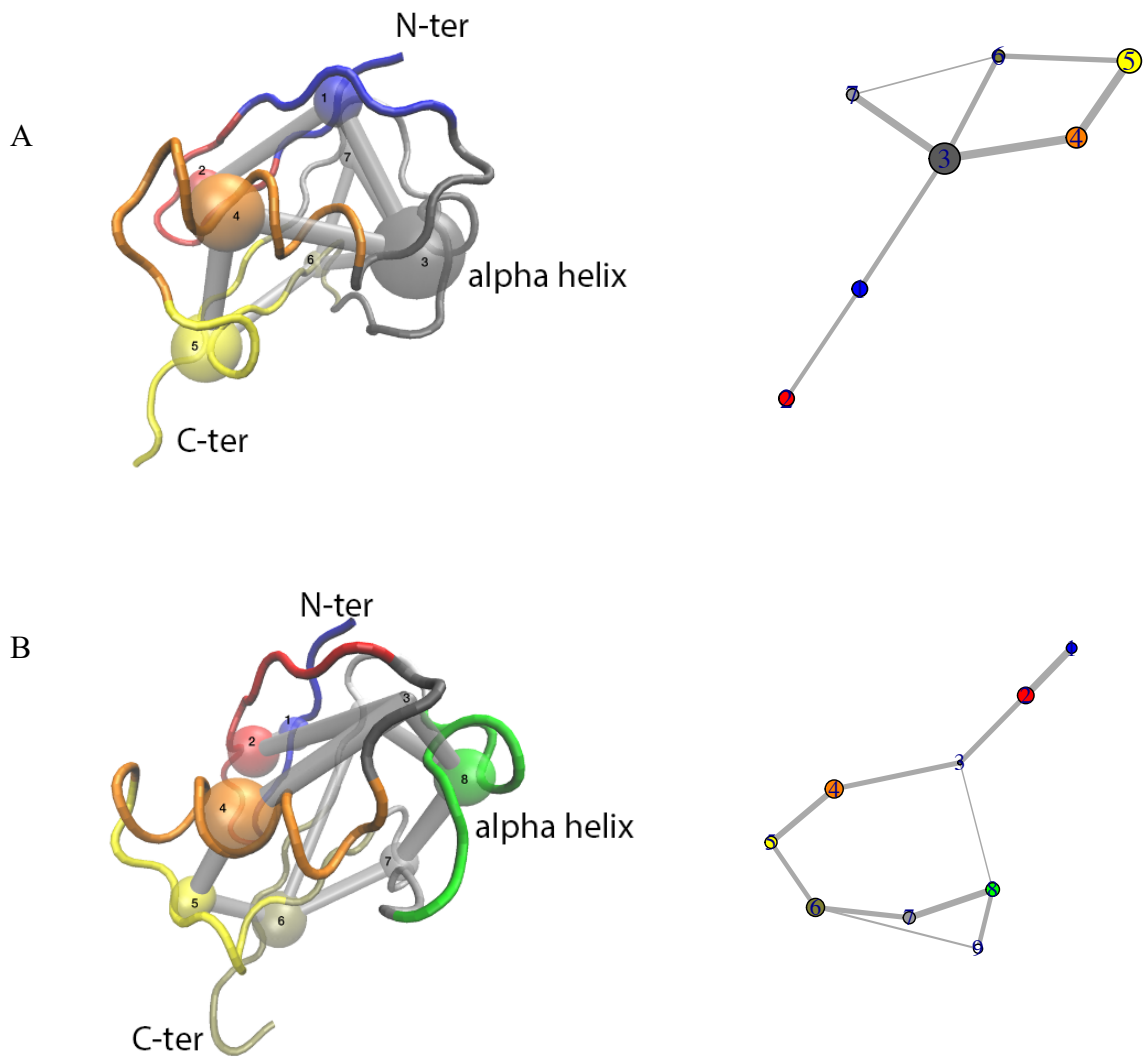

Figure 6. Community network analysis of ubiquitin dynamics. On the left, the colored nodes are superposed on the protein backbone structure, represented as a tube and colored according to the nodes. The gray edges between nodes represent pathways between two nodes, where the thickness of the edge indicates the strength of the correlation. On the right is the 2D network representation. In (A) are the results from the AA simulation, and in (B) are the results from the Drude-2019 simulations.

The results from the community network analysis of ubiquitin using the correlated motions from the AA simulation are shown in Fig. 6A. The specific compositions of the nodes are given in Table S1 of Supporting Information. The analysis of the AA simulation of ubiquitin resulted in 7 community nodes and 8 edges. The most prominent node puts residues of loop S2 – H1 and residues of H2 in the same community node, N3, a large node containing 18 residues. These regions show a high degree of correlated motion between them and are therefore grouped together in a single node (Fig. 6A). Node 3 also has connections to several other nodes suggesting that this region is highly correlated to much of the protein. These connections between communities are also seen in the correlation map (Fig. 4), but the analysis here highlights better the topological features of ubiquitin. Correlations are identified along secondary structure elements, emphasizing the critical position of the loop S2-H1 and H2 region in the protein, as it shows correlations with regions shown to exhibit the pincer motion[17,18].

The CNA analysis of the Drude-2019 simulation of ubiquitin gives 9 community nodes and 10 edges (Fig. 6B). We see practically all of the secondary structure elements having their own nodes, and this higher number of nodes suggests that there is more decoupled motion in the simulation of this protein. Node 3, which encompassed several structural elements in the AA simulation, now contains only 3 residues (residues 19:21 between S2 and H1). In the AA simulation, this region showed much stronger correlations. This significant change in the calculated network architecture is most likely due to the smaller values of the correlations in the Drude-2019 simulations (see Fig. 4).

**PPARγ**

*Drude-2019 FF simulation shows larger RMSD time series values*

The RMSD time series for PPARγ were calculated from the molecular dynamics simulations for the three replica of the AA and Drude-2019 simulations for each system. For each system, the average times series of the three replicas were displayed along with the high/low values at each time point. All four PPARγ systems show stable 100 ns trajectories (Fig. S4). The RMSD mean value of PPARγ-apo system simulated with the Drude-2019 model was higher than the value of the system simulated with the AA force field with the values of 3 Å (SD: 0.09) and 2.5 Å (SD: 0.06), respectively (Fig. S4 A,B).

We notice the same trend when comparing the simulations of PPARγ bound to the corepressor peptide NCoR (Fig. S4 C,D). The Drude-2019 simulations presented higher values of RMSD, with the mean value of 3.2 Å (SD: 0.21) than the AA simulations, where the mean value is 2.4 Å (SD: 0.04). These results are consistent with the conclusions that the Drude-2019 force field allows for a higher conformational flexibility than the standard additive CHARMM force field[7]. In addition to the overall stability of the PPARγ-corepressor bound complex, we see the interaction of two components being stable as the peptide does not dissociate from the PPARγ LBD, confirming that the Drude-2019 force field maintains well protein-peptide complexes.

*Drude-2019 force field simulations show more significant RMS fluctuations*

We calculated the RMSF of the backbone atoms of PPARγ-apo averaged by residue over all three replicas. In all of the cases, we observe an RMSF profile that reflects the stability of the structure, where loops are more flexible than the secondary structure regions. While the AA simulations present the highest flexibility in the regions of the loop between H2 – S1, loop H9 -

H10 and H12 (Fig. 7A), the Drude-2019 simulation shows a higher flexibility in the loop H2-S1, the ω loop region (between H2' and H3) and smaller flexibility of the loop H9 - H10 (Fig. 7B). This lower flexibility of loop H9-H10 coincides with a salt bridge between residues D411 (H8) and H453 (loop H9-10). This salt bridge is generally conserved in crystal structures of class II NRs[47], and it was maintained in the simulations with the Drude-2019 force field, but was mostly lost in the AA simulations. The second salt bridge characteristic of class II NRs, between residues E352 (mid H4 – 5) and R425 (loop H8 – 9), is well maintained in both AA and Drude-2019 simulations. Time series for the D411-H453 salt bridge is given in Supporting Information, Figs. S5A and S5B, for apo PPARγ and PPARγ complexed with the corepressor peptide, respectively.

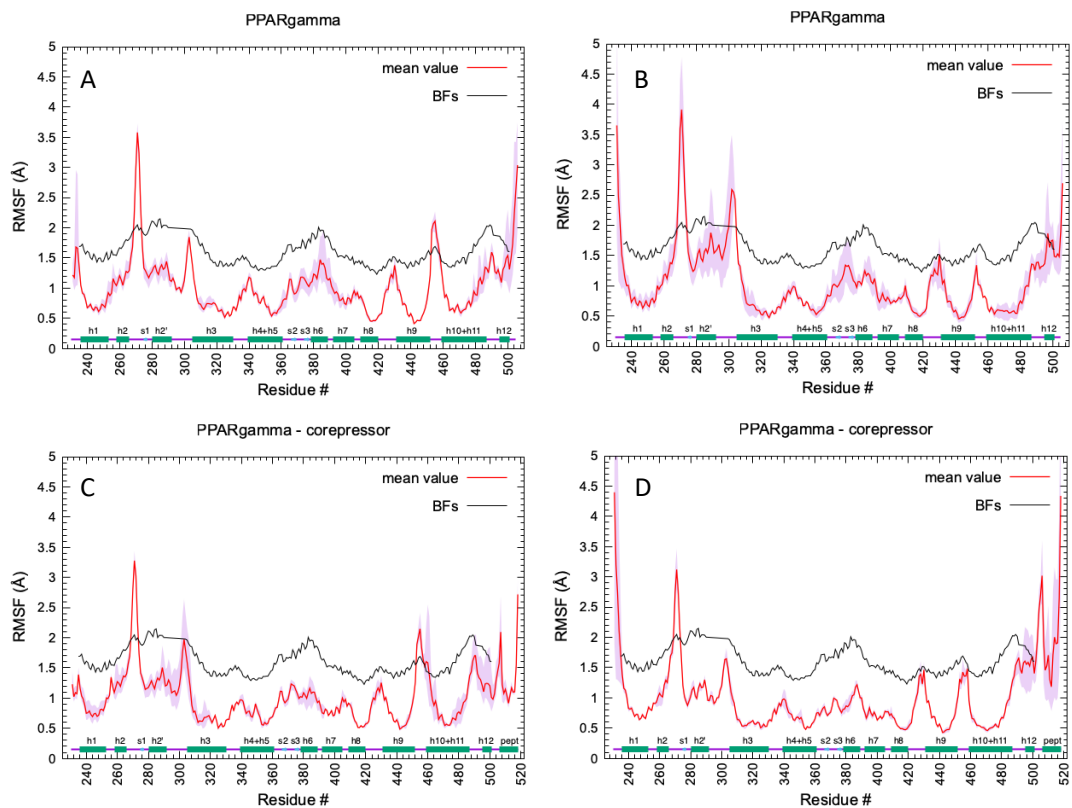

Figure 7. RMSF of PPARγ LBD by (A) AA simulations, (B) PPARγ LBD by Drude simulation. PPARγ LBD bound to the corepressor peptide by (C) AA simulation, (D) PPARγ LBD bound to the corepressor peptide by Drude simulation. The mean value of 3 replicas is represented as a red line. Secondary structure elements are shown on the x axis: alpha helices (H1 – H12) as green, and beta strands (S1 – S3) as blue rectangles. The black BFs line show the rms fluctuations calculated from crystallographic B factors.

For the PPARγ - NCoR system, the differences are less prominent, the RMSF curves for both the AA and Drude-2019 simulations are similar, albeit with differences in the H9 - H10 loop and the H12 and corepressor peptide region (Fig. 7C,D). Higher variability is found in the AA simulation around H2' and the ω loop, and also in the loop H9 - H10. Comparing the apo and corepressor bound PPARγ systems, we see the difference in the β-sheet region and H6. For both AA and Drude-2019 simulations, adding the corepressor peptide lowered the replica-averaged RMSF. With the Drude-2019 simulations, the variability among replicas is also much smaller. As all the PPARγ structures simulated originated from the same crystal structure, the fluctuations from B-factors are the same in all the plots. Comparison with the fluctuations calculated from the simulations are in good agreement with the fluctuations from experimental B-factors as they all show the same trends. Concerning the D411-H453 salt bridge, the same result was found here as for the apo structure - the salt bridge is maintained in the Drude-2019 simulation but essentially lost, appearing for short times in two out of three replicas of the AA simulations, see Figs. S5B.

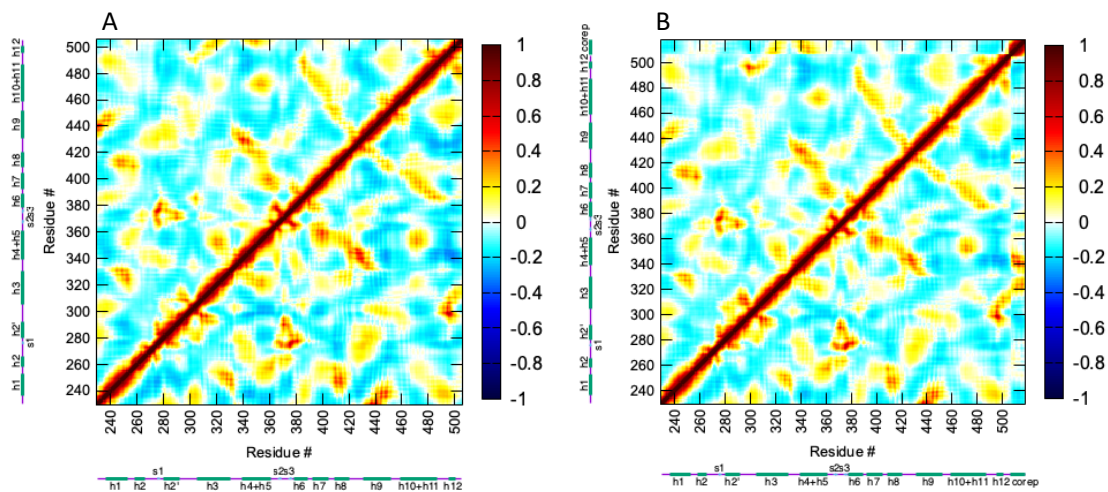

Figure 8. Correlated motions calculated from the simulation of the PPARγ LBD apo form (A), and corepressor-bound form (B), comparing simulations from the Drude-2019 simulation (upper triangle) to the AA simulations (lower triangle). Maps are represented with a color code related to the sign and intensity of correlations (ranging from dark blue for perfect anticorrelations to dark red for perfect correlations). The secondary structure elements are indicated.

We assessed the effects of polarization on the correlated motions from the MD simulations of the PPARγ LBD apo form by calculating the cross-correlations, as done for ubiquitin (see Fig. 8). The upper triangle of the map corresponds to the correlated motions of PPARγ calculated from the Drude-2019 simulations, while the lower triangle corresponds to the correlation map for PPARγ calculated from the simulation using the AA force field. The calculations were done for both the apo and corepressor-bound forms. Regarding the general aspect of the correlation maps, for both forms, we notice a great similarity between the two force fields. The differences are noticeable regarding correlation intensities, as they are lower in the Drude-2019 simulations, to the point that, in particular regions, correlation islands disappear. In the case of the PPARγ-apo system, the most significant isles represent the correlation between H12 - ω loop residues, and H1-H9. While strongly present in the AA maps, only traces of the isles are present in the Drude-2019 simulations.

A scatterplot of the calculated correlation coefficients of PPARγ from the AA and Drude-2019 simulations shows an overall linear correlation between the two sets of data, but further shows that the Drude simulation data are lower (in absolute value) than the AA simulation data (Fig. 9A). In Fig. 9B, the probability distribution of the correlation coefficients are plotted for the AA and the Drude-2019 simulations. The distribution is broader for the AA simulations and extends further to negative values (anticorrelated motions) for the AA simulations. A visual inspection of the data indicates that particularly large discrepancies (0.3 and above; 0.5% of the data) correspond to correlations made between H12 and the C-ter end of the ω loop and N-ter of H3, a region of functional importance. In the AA simulations, there are clear correlations between these two regions which essentially disappear in the Drude-2019 simulations. Note, that in the RMS fluctuations, there was a difference between the two force fields in the ω loop region as well (Fig. 7).

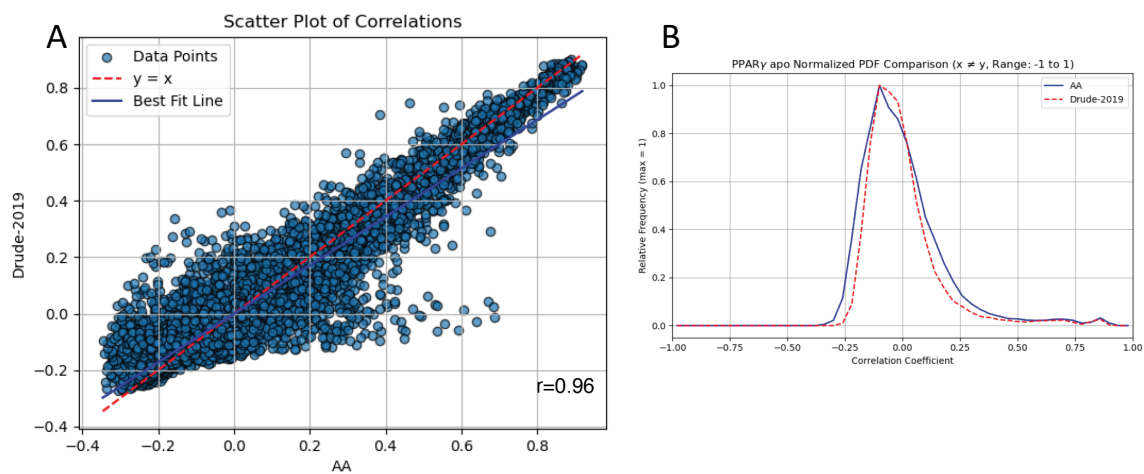

Figure 9. (A) Scatterplot of the correlation coefficients used to calculate the correlation maps from the molecular dynamics simulation of PPARγ in apo form; from the AA simulations (x-axis) vs from the Drude-2019 simulations (y axis). The dashed line shown in red is x=y; the solid blue line is the best fit to the data excluding the self-correlation coefficients which equal 1, (B) the probability distribution of the correlation coefficients from the AA simulation (solid blue) and the Drude-2019 simulation (dashed red)

PPARγ bound to the corepressor is the system where the correlation between the AA and Drude data is the strongest (Fig. S8 in Supporting Information) with a few outliers in the scatterplot. A few correlations differ by 0.2 or more (about 0.5% of the data). These outliers are generally found between amino acids that are in proximity and solvent exposed, for example between helices H1 and H9, helix H2' and the C-ter of H5, between helices H12 and H3, and between H12 and H4. Helices H3, H4 and H12 constitute the platform for corepressor binding. The probability distribution of the correlation coefficients show a similar trend as observed for the PPARγ-apo (Fig. S8B), but the differences are less significant.

*CNA analysis of PPARγ*

Community network analyses of the PPARγ were carried out using the correlations calculated from the molecular dynamics simulations. The results for PPARγ-apo are shown in Fig. 10, and the results for PPARγ with corepressor peptide are given in Supporting Information Fig. S9. The specific compositions of the nodes are given in Supporting Information, Table S2.

Differences in calculated correlations discussed in the context of the correlated motions maps (Fig. 8) lead to the community network analysis of the PPARγ differing in an important manner. These results lead to different interpretations relating to the functional role of the two PPARγ forms, depending on the force field used.

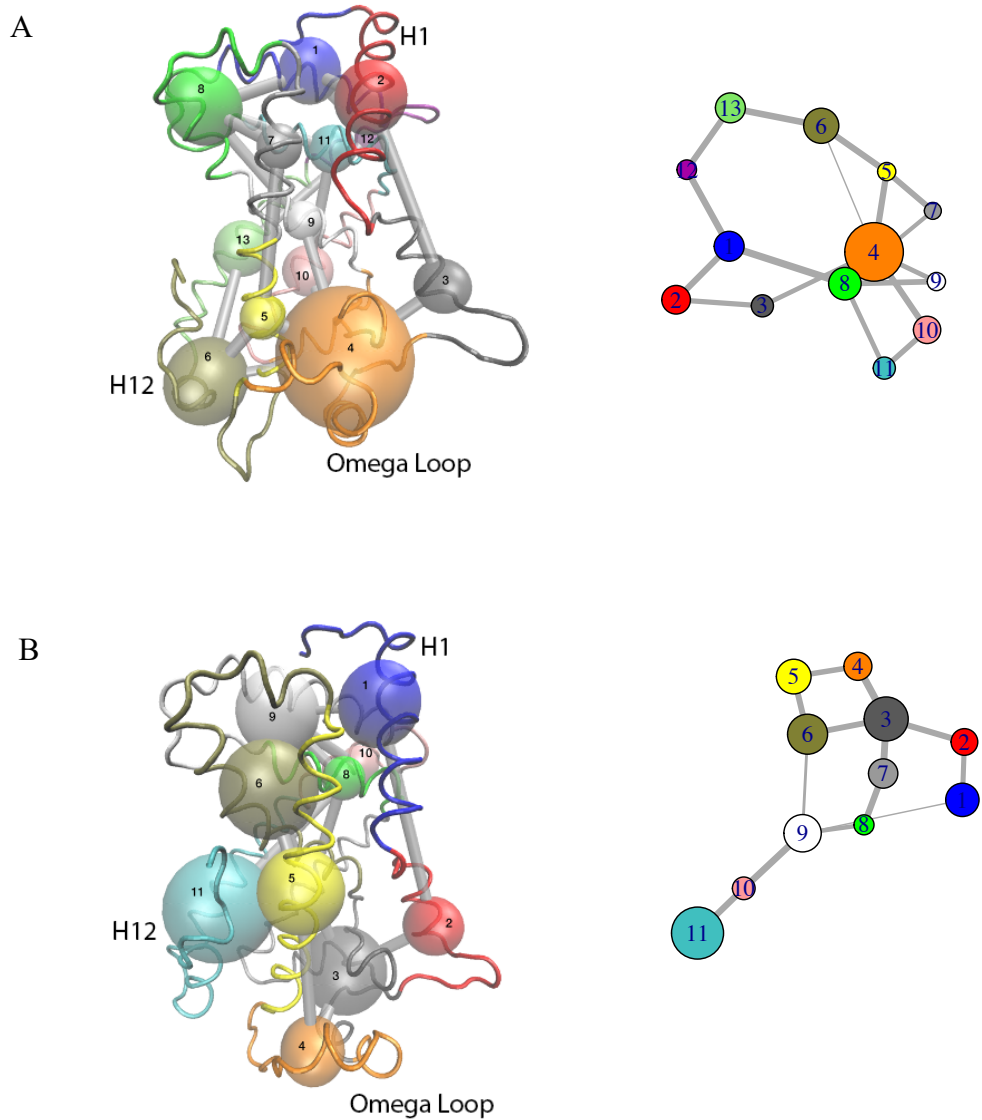

Figure 10. Community network analysis of the PPARγ LBD dynamics. On the left, the colored nodes are superposed on the protein backbone structure, represented as a tube and colored according to the nodes. The gray edges between nodes represent pathways between two nodes, where the thickness of the edge indicates the strength of the correlation. On the right is the 2D network representation. In (A) are the results from the AA simulation, and in (B) are the results from the Drude-2019 simulations.

In particular, the CNA shows that many of the node interconnections are along the secondary structure elements, but the AA simulations display several edges between nodes beyond secondary structure (see Fig. 10). The node organization for AA is more spatially broad than for the Drude-2019 simulation results, where there is less interconnectivity and the network is more extended. Central to the interconnectivity in PPARγ is the node that encompasses the β sheet and part of the ω loop region. This node forms a hub through which many edges connect in the AA, but not in the Drude-2019 simulations.

One significant distinction between the simulations using the two different force fields concerns the community that represents helix H12. In the AA simulations of both apo and corepressor-bound systems, H12 and part of the ω loop are coupled and are therefore represented by one community. In the Drude-2019 simulations, H12, together with H11 make up an individual community in both structures. This community is decoupled from the node encompassing the ω loop in both systems simulated by the Drude-2019 FF meaning there are no edge connections between them. This suggests that the correlations in the Drude-2019 simulations are not sufficiently strong to result in the CNA analysis detecting direct communication between H12 and the ω loop region. In the corepressor-bound form, H12 is further decoupled from H11, having its own community of helix residues connected by an edge to the H11 (Fig. S9 in Supporting Information).

Helix H12 represents the Activation Function 2 (AF-2) in LBDs and therefore is physiologically important for the regulation of PPARγ's transcriptional activity. In the community network analysis of the Drude-2019 simulations, we noticed the decoupling of the H12 from the other regions, notably the ω loop and the H11. This suggests that these regions explore different movements which are not directly correlated and display different conformational dynamics.

Unlike ubiquitin, there are few experimental studies at the atomic level that specifically studied correlated motions in PPARγ. Nevertheless, there is substantial evidence from global studies of its dynamics and from physiological studies that underline the presence of functional motions and their role in communication between different regions of the nuclear receptor ligand binding domain. For example, from experimental studies, it is known that phosphorylation of Ser273, located in the ω loop region of the PPARγ ligand-binding domain significantly alters interactions between the LBD and coregulator proteins, even though the phosphorylation site and coregulator interaction sites are distant. The phosphorylation of Ser273 leads to changes in gene expression, particularly in metabolic and insulin-related pathways[48–50]. Ser273 phosphorylation reduces the affinity of PPARγ for coactivators like PGC-1α and SRC-1, which leads to decreased transcriptional activation of insulin-sensitizing genes such as adiponectin and Glut4. The work of Gonçalves Dias *et al* showed that the phosphorylated form of PPARγ had an increased interaction with corepressors like NCoR (Nuclear Receptor Corepressor) and SMRT (Silencing Mediator for Retinoid and Thyroid Receptors)[50]. The principal helices involved in coregulator binding are H3, H4 and H12. We see from the CNA that in the AA simulations, these helices show interconnectivity of nodes of correlated amino acids, while in the Drude-2019 simulations, these important regions and, significantly, the transcriptionally important H12 show no apparent connectivity to the ωloop which contains the Ser273 phosphorylation site.

The ω loop is an important hub for PPARγ activity, as highlighted by the effect of point mutations on transcriptional activity. The mutations I267A and F287A abolish prostaglandin activation of PPARγ[51]. The point mutations M280I, I290M of PPARγ, identified in bladder tumors, gave proteins with a significantly higher levels of transcriptional activity than the wild type (WT) in the absence of exogenous ligand (two to six times higher)[52]. M280I and I290M are also located in the ω loop of the LBD.

These results suggest that modifications in the ω loop can mediate activation of PPARγ through allosteric communication with H12. In the AA-simulations, the CNA results show that the ω loop consists of a large, well-connected node N4, while in the Drude-2019 calculations, the node is much smaller and less-well connected.

Another significant distinction between the simulations using the two different force fields concerns the community that represents helix H12. In the AA simulations of both apo and corepressor-bound systems, H12 and part of the ω loop are coupled and are therefore represented by one community. As the overall correlation between the AA and Drude correlation coefficients is good (see Figures 5, 9 and Fig. S8 in Supporting Information), we checked if lowering the default threshold for the CNA analysis specifically for the Drude simulations (for the PPARγ-apo system) would bring the CNA conclusions of the two force fields in line for this physiologically important region. As presented in Supporting Information Section 3, this simple change was not sufficient to bring the two force fields in agreement.

**Hydrophobic cluster analysis**

Advanced analysis of correlation networks by methods such as community network analysis (see above), as well as by also Shortest Path method (see Supporting Information Sec. 3), lead to markedly different interpretation of allosteric communication between the AA and Drude simulations. This is presumably due to the weaker correlated motions. In an effort to better understand why the correlated motions are weaker in the Drude simulation, we examined the hydrophobic contacts in ubiquitin and in PPARγ. NMR studies have suggested that hydrophobic clusters in proteins provide anchor points for long range collective motions. In the work by Bouvignies *et al*[53], they identified slow correlated motions that implicated hydrophobic sidechains buried in the core of an Ig-binding domain of streptococcal protein G. Kim *et al*[54] also found by NMR that the hydrophobic core of the catalytic subunit of protein kinase A moved in a correlated manner in response to adenosine 5′-triphosphate binding, revealing a correlated hydrophobic network. Hydrophobic cores have been associated with allosteric motions in other proteins as well[55,56].

Using the Protein Tools webserver, we analyzed the ubiquitin and PPARγ hydrophobic clusters based on Ile, Leu and Val. It has been advanced that ILV clusters prevent the intrusion of water molecules and serve as cores of stability in high-energy partially folded protein states[57]. The clusters determined by the Protein Tools webserver are defined in Supporting Information, Section 4. For the analysis of the trajectories, we used the largest clusters, which are shown in Fig. 11.

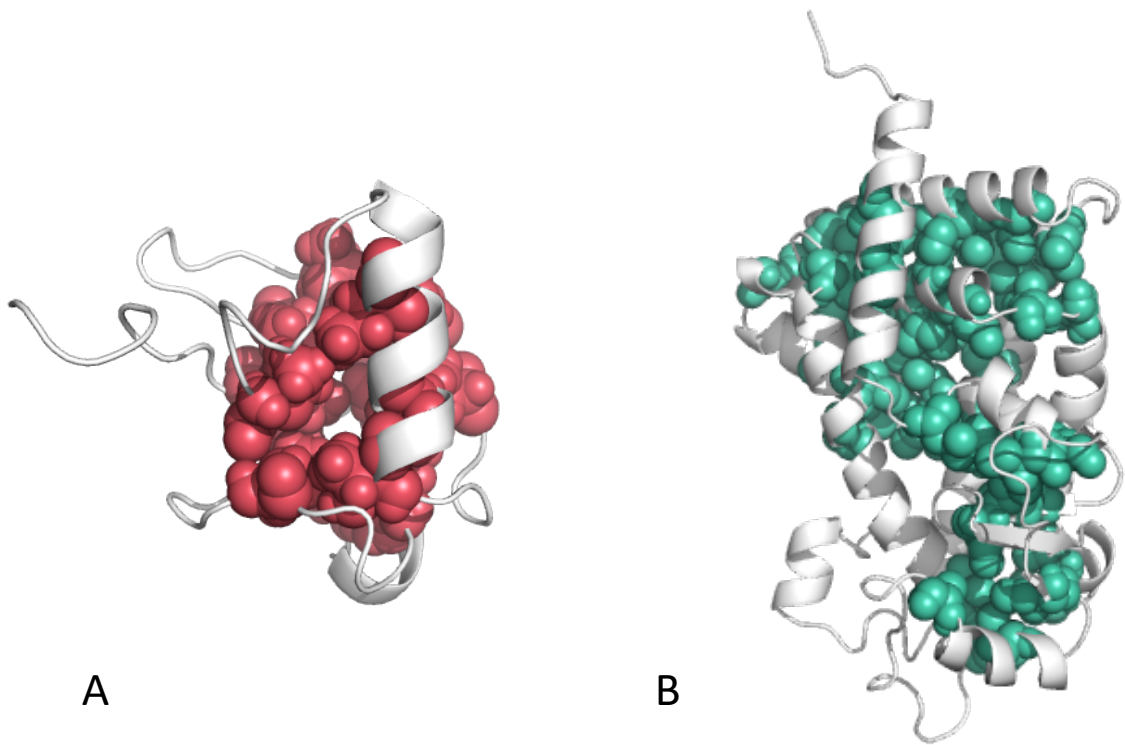

Figure 11. Hydrophobic cluster determined by the ProteinTools webserver used in the analysis of the molecular dynamics trajectories. A) Ubiquitin, B) PPARγ-apo

As the clusters were determined using the starting structures for our simulations, we took these definitions as being the native contacts. Using the CHARMM program, we then analyzed the trajectories to calculate Q, the fraction of native contacts observed in each frame of the trajectory during the trajectories. The results are given as box plots in Fig. 12.

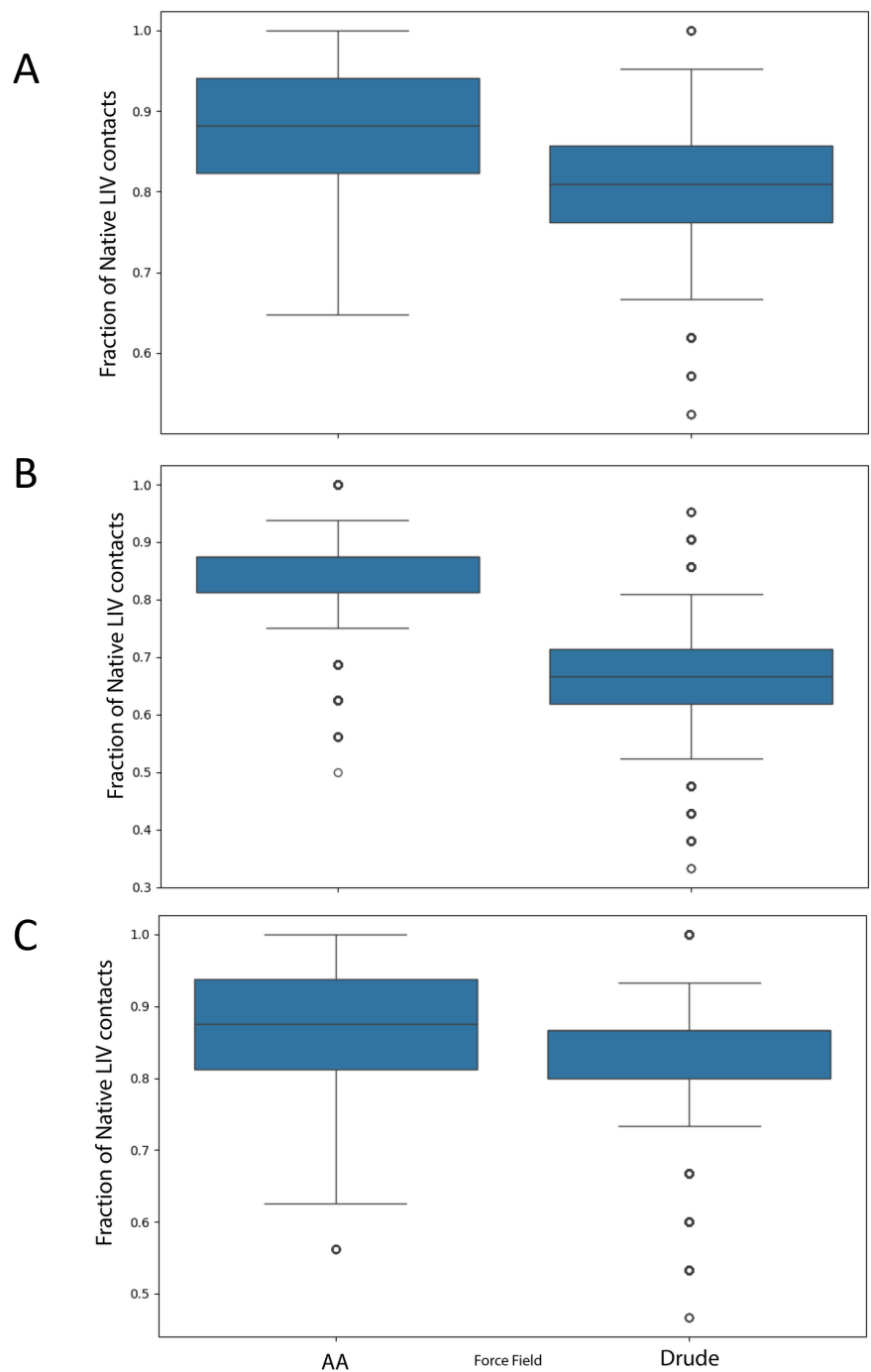

Figure 12. Box plots of native LIV-contacts calculated from the molecular dynamics simulations. A. Ubiquitin, B. PPARγ-apo, C. PPARγ-corep

From Fig. 12, we see a systematic loss of native hydrophobic contacts involving L, I and V in the Drude-2019 simulations. A Wilcoxon rank-sum test showed that the differences in number of contacts are statistically significant and that the data sets are statistically independent (p values =0, for complete statistical analysis results see Supporting Information Section 5). An energy analysis of the clusters was done and is provided in Supporting Information.

These data suggests that the disruption of native hydrophobic contacts can have an effect on the long range correlated motions. Between the analysis of the hydrophobic cluster contacts and the energetic analysis (Sec. 6 in Supporting Information), our results suggest that the Drude-2019 model may underestimate the strength of van der Waals interaction in hydrophobic clusters of amino acids, which in turn would impact long-range correlated motions and their related interpretations that aim to identify allosteric networks.

**Conclusions**

In this work, we used the Drude-2019 polarizable force field in molecular dynamics simulations of two proteins, human ubiquitin and the ligand binding domain of human PPARγ. We compared the results to simulations using the CHARMM all atom force field. We examined the effect of explicit polarization on standard measures of structural dynamics, such as RMSD and RMSF. We generally found conformational changes leading to a higher RMSD and, in flexible regions of the proteins, greater flexibility when using the Drude-2019 force field.

We also characterized for the first time the effects of using the Drude-2019 force field on correlated motions, which are implicated in the biological function of proteins. The correlated motions were characterized by correlation maps calculated from molecular dynamic simulations and further analyzed by community network analysis (CNA). The CNA identifies regions of the protein where residues are strongly correlated in their motions; these are grouped into communities (nodes). Longer-range correlations between the nodes are identified by connections between the nodes. The analysis can reveal paths through which signals can propagate from one region to another and thus suggest molecular mechanisms of allosteric communication. Analysis of the correlated motions showed significant difference between the two force fields, with the Drude-2019 simulations yielding overall weaker, decoupled correlations. This in turn had a significant impact on the CNA analysis.

A comparison of the CNA results with available experimental data on allosteric communication in the two proteins studied indicates that the CHARMM all atom additive force field results are in better agreement with experimental observations. The reasons for this can be linked to several factors. Indeed, even though the polarizable force field represents a more physically correct description of interatomic interactions with proteins, it still contains several approximations with respect to a full quantum mechanical description. The development of non-polarizable force fields over the years has led to force fields that manage a delicate balance of forces and are in agreement with a large and diverse amount of experimental data. Iterative refinement over several years of testing against experimental data has led to the current CHARMM all atom force field. The Drude-2019 force field has undergone several iterations of development from the 2013 version[7] to the 2019 version[9] with the latter being distributed in the Drude Prepper application of CHARMM-GUI platform[19]. Testing in new systems and against diverse data is still necessary to achieve a fully balanced description. Indeed, shortcomings of the Drude-2019 force field were recently discussed in modelling of conformational equilibria of complex systems[58–60]. These results are in line with other studies of physiological processes where hydrophobic contacts are important, for example in protein aggregation[61] and protein folding[62] where similar conclusions that the Drude-2019 force field does not lead to improved results were reached.

Our analysis suggests that the weaker correlations evident in the correlation maps are likely due to the decrease in native hydrophobic contacts, which are crucial anchors for long-range correlated motions and depend on the van der Waals interactions modelled by a Lennard-Jones potential in the CHARMM force fields. The necessity to further optimize the Lennard-Jones parameters in the context of the Drude-2019 model has been pointed out by the developers of the force field[63]. These weaker correlations led to CNA maps that were more difficult to interpret in the context of what is known about the dynamics (ubiquitin) and physiological behavior of the proteins (PPARγ). It should be pointed out that the development of correlated motions analysis tools such as CNA was based on additive force field simulations, and therefore, the recommended threshold for correlation coefficients may not be optimal in the context of simulations using the Drude-2019 force field. Lowering the recommended threshold of the CNA analysis specifically for the Drude-2019 simulations did not lead to communities and edges in better agreement with the additive force field data; simple modulation of the threshold was not sufficient to bring the two analyses in agreement. This question of threshold values in CNA analysis warrants future investigation. So, while it has been shown that polarization can enhance local cooperative folding of an alpha helix [64], when considering long-range correlated motions where hydrophobic interactions are important, our results advise use of the classical additive CHARMM36m force field.

## ASSOCIATED CONTENT

**Supporting Information**

A file (pdf) containing additional text, figures and tables for

- Program and Data Availability
- RMSD time series for ubiquitin, apo-PPARγ and PPARγ-corepressor complex
- Electric dipole moments analysis from the AA and Drude-2019 simulations of ubiquitin and PPARγ
- Timeseries for the D411-H453 salt bridge for PPARγ-apo and PPARγ-corepressor complex
- Scatterplot of the correlation coefficients for PPARγ corepressor peptide complex
- Table listing the Composition of the nodes from the community network analysis of ubiquitin
- Table listing the Composition of the nodes from the community network analysis of PPARγ
- Detailed discussion of the CNA analysis for the PPARγ LBD
- Shortest Path Method analysis of correlated motions
- Hydrophobic ILV cluster analysis with listing the hydrophobic cluster composition determined by the Protein Tools webserver
- Detailed results of the statistical analysis of the native hydrophobic contacts and the self van der Waals and electrostatics energies

AUTHOR INFORMATION

**Corresponding Author**

**Roland H. Stote**, *Institute of Genetics and Molecular and Cellular Biology, 1 Rue Laurent Fries, 67400 Illkirch-Graffenstaden, France,* orcid.org/0000-0003-2230-7145

Email: rstote@igbmc.fr

**Authors**

Ana Milinski*, Institute of Genetics and Molecular and Cellular Biology, 1 Rue Laurent Fries, 67400 Illkirch-Graffenstaden, France,* orcid.org/0009-0004-4530-6350

Annick Dejaegere, *Institute of Genetics and Molecular and Cellular Biology, 1 Rue Laurent Fries, 67400 Illkirch-Graffenstaden, France*, orcid.org/0000-0002-3569-7771

**Author Contributions**

The computer studies were conceived and executed by AM and RHS. All authors contributed to the analysis and the manuscript was written with contributions from all authors. All authors approved the final version of the manuscript.

**Funding Sources**

This work was supported by the *Agence Nationale pour la Recherche* (FRIDaY project ANR-20-CE29-0013) for financial support. Computational resources were provided by Strasbourg University High Performance Computing Center and GENCI (Grand Equipement National de Calcul Intensif).

**Notes**

The authors declare no competing financial interest.

**Acknowledgements**

The authors acknowledge the *Agence Nationale pour la Recherche* (FRIDaY project ANR-20-CE29-0013) for financial support. The computational work was supported by the Centre National de la Recherche Scientifique, the Institut National de la Santé et de la Recherche Médicale and the University of Strasbourg. The authors acknowledge particularly the support of the Strasbourg University High Performance Computing Center and GENCI (Grand Equipement National de Calcul Intensif) for computing resources. AM would like to thank Dr. Nastazia Lesgidou of the Biomedical Research Foundation Academy of Athens, Greece, for discussions on Community Network Analysis. Clara C. Stote is acknowledged for help with the illustrations. The authors would like to thank Meilin An and Valentin Loux for helpful discussions and reading of the manuscript.

ABBREVIATIONS
LBD, Ligand binding domain; PPARγ, Peroxisome Proliferator-Activated Receptor gamma

TOC Graphic

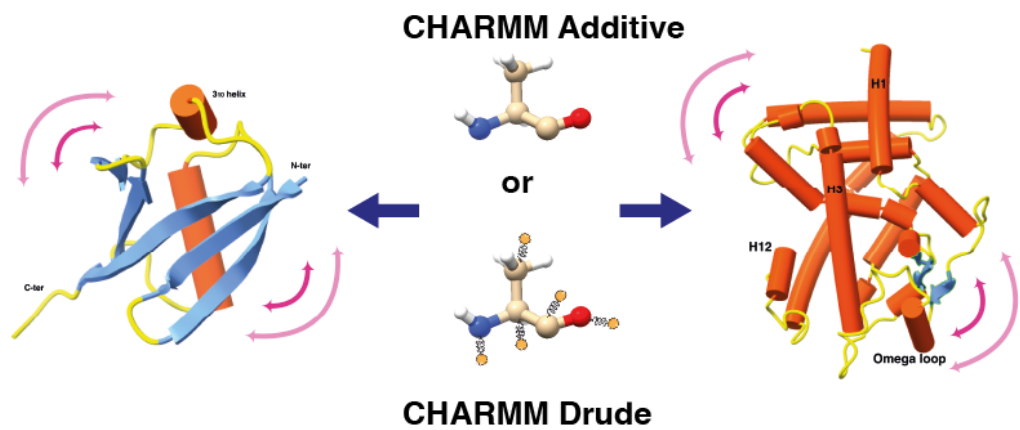

Supporting Information For

# Impact Of Force Field Polarization On Correlated Motions Of Proteins

Ana Milinski, Annick Dejaegere, Roland H. Stote

Institute of Genetics and Molecular and Cellular Biology, 1 Rue Laurent Fries, 67400 Illkirch-Graffenstaden, France

**Program and Data Availability**

The NAMD[1] and CHARMM[2] molecular modeling simulation programs used in this study are available at https://www.ks.uiuc.edu/Research/namd and https://brooks.chem.lsa.umich.edu/register/, respectively. The VMD visualization program[3] (version 1.9.4) is available from https://www.ks.uiuc.edu/Research/vmd/. CHARMM-GUI[4] is accessible free of charge for academic users at https://www.charmm-gui.org/. The R package[5] and Bio3D[6] are available at https://www.r-project.org/ and http://thegrantlab.org/bio3d/, respectively. The PROPKA program was accessed through the website https://server.poissonboltzmann.org/pdb2pqr in Sept 2021. The Protein Tools[7,8] (accessed in spring of 2025) can be found at https://proteintools.uni-bayreuth.de/clusters/documentation. The Shortest Path Method (SPM)[9,10] (version available in 2024) is available through the online webserver https://spmosuna.com upon request to the webserver authors. All initial structures, sample simulation scripts, parameter and topology files are shared through a link hosted on Zenodo: https://zenodo.org/records/14291250. Additional analysis scripts are available upon request to the authors.

**Section 1: RMSD time series and additional analysis**

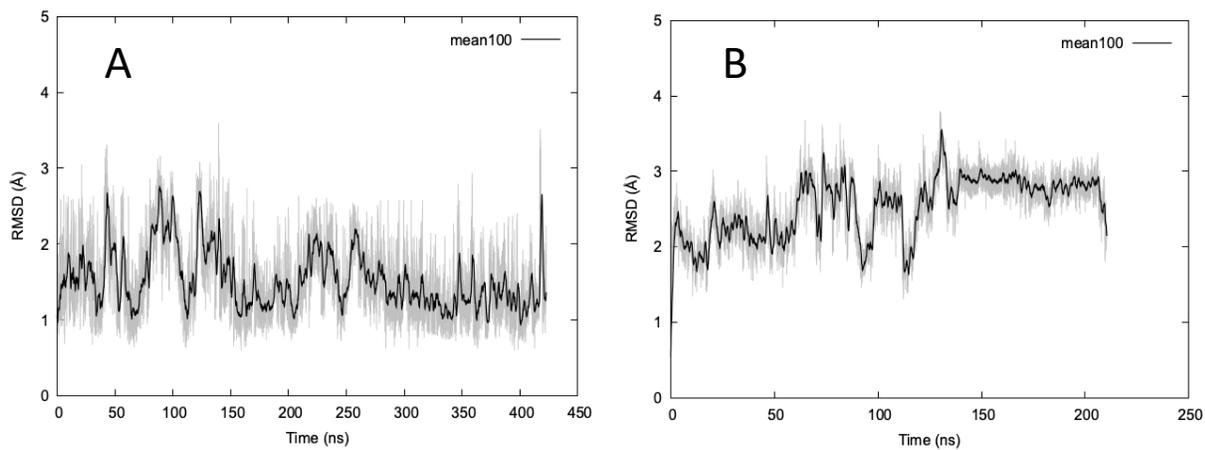

Figure S1: RMSD of ubiquitin backbone atoms with respect to the initial structure. Results from the simulation using the AA force field (A) and the simulation using the Drude-2019 force field (B) are shown in gray, black lines correspond to the running average over 100 data points.

**Electric dipole moments from the AA and Drude-2019 simulations**

Electric dipole moments of proteins contribute to the structural dynamics and function by influencing how proteins interact with their environment. Accurate modeling of the dipole moment can thus improve the overall representation of intermolecular interactions. The dipole moments of ubiquitin and PPARγ were calculated along the trajectories. The data are presented as time series of the dipole moment of the full protein. We further calculated the average dipole moment of the protein backbone by-residue.

For ubiquitin, the comparison of the dipole time series calculated from the AA and the Drude-2019 simulations shows that, overall, both force fields give average values and fluctuations with roughly the same magnitude (Fig. S1). The by-residue analysis shows that the amino acids in alpha helices systematically show larger backbone dipole moments in the Drude-2019 model than in the AA force field (Fig. S2), similar observations were made by Lopes *et al* [11]. In beta sheets, the opposite is generally observed, that the backbone dipole moments calculated for the Drude-2019 simulations are lower than those in the AA simulations.

We calculated dipole moment timeseries from the AA and Drude-2019 simulations of the apo PPARγ protein and the apo PPARγ protein complexed to the corepressor peptide (Fig S3). For the PPARγ apo system, we see lower dipole values in simulations with AA force field, with the average of 247 D, compared to the Drude-2019 force field, where the average value is 329 D. The calculations of PPARγ corepressor bound system follow the same trend, where the average values are 240D and 332D for the AA and the Drude-2019 FF, respectively.

Concerning the by-residue average dipole moment, we see that in the case of PPARγ, the variation of residue dipole moments is much more significant than what was observed in ubiquitin for both the apo protein and the protein complexed with the corepressor peptide (Fig. S4). Systematically, the dipole moments in alpha helices are larger in the Drude-2019 simulations than in the AA force field simulations for PPARγ.

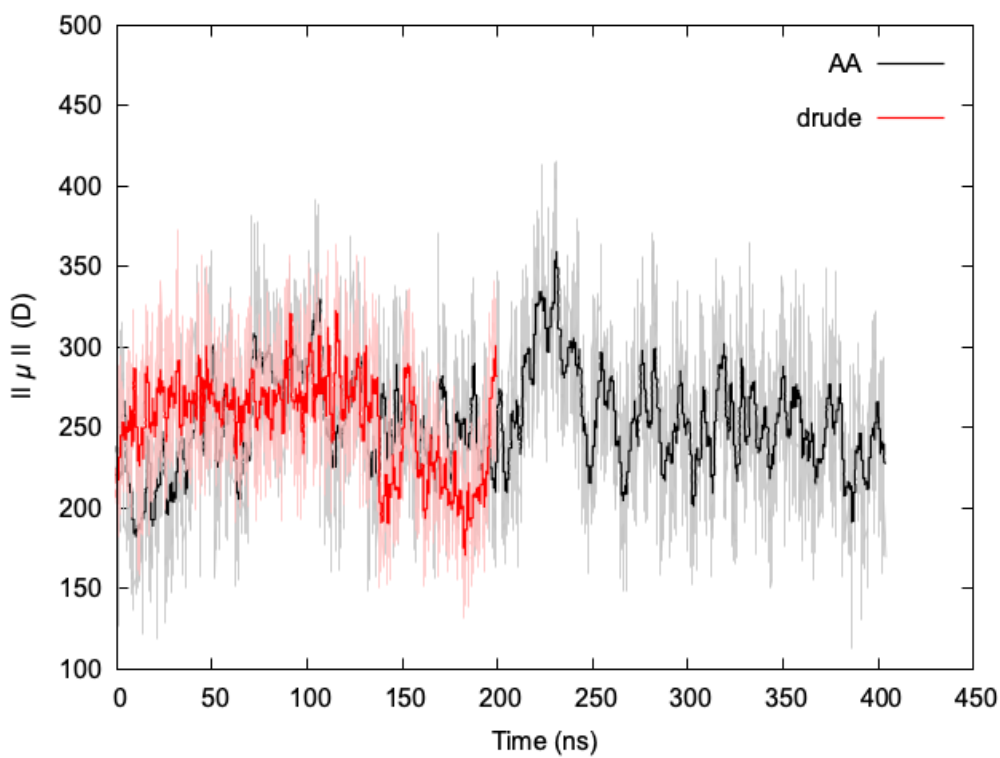

Figure S2. Time series for full dipole of ubiquitin. Shown is the running average over 100 frames in black (AA) and red (Drude-2019) along with shades of individual values during the time series.

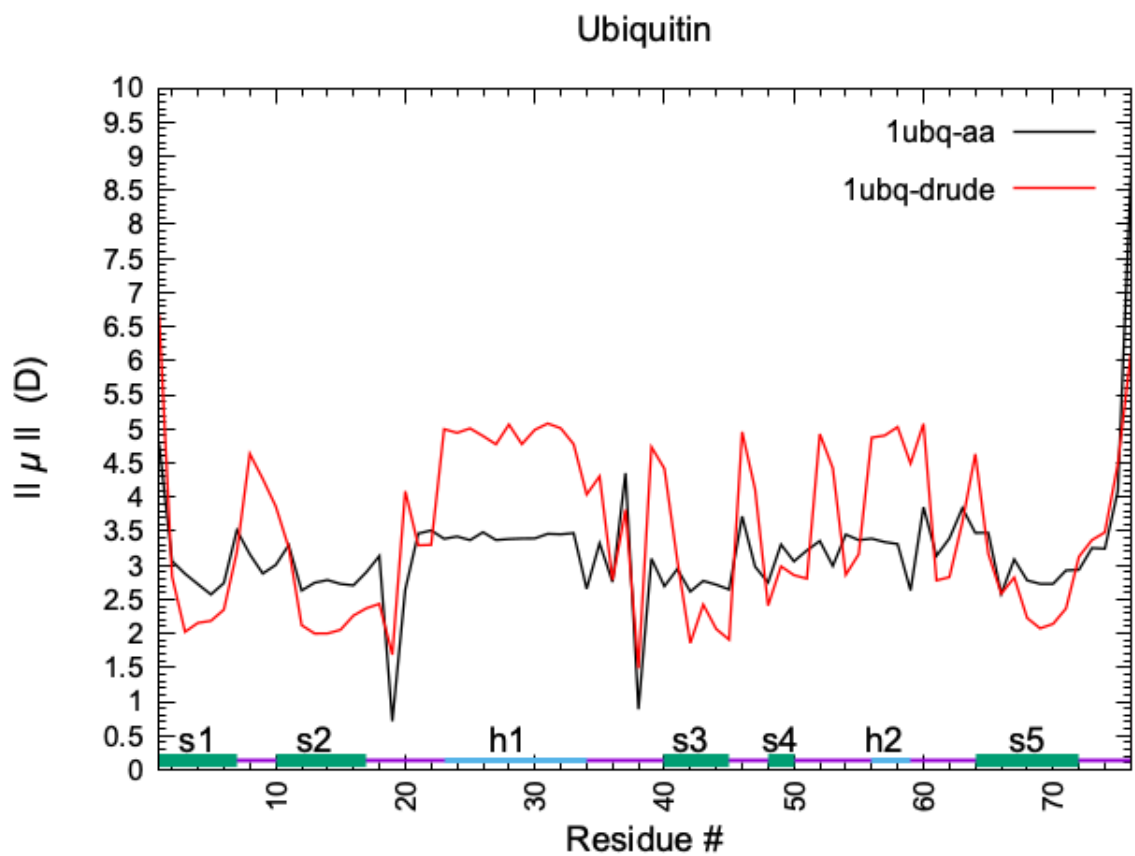

Figure S3. Average dipole moment by-residue.

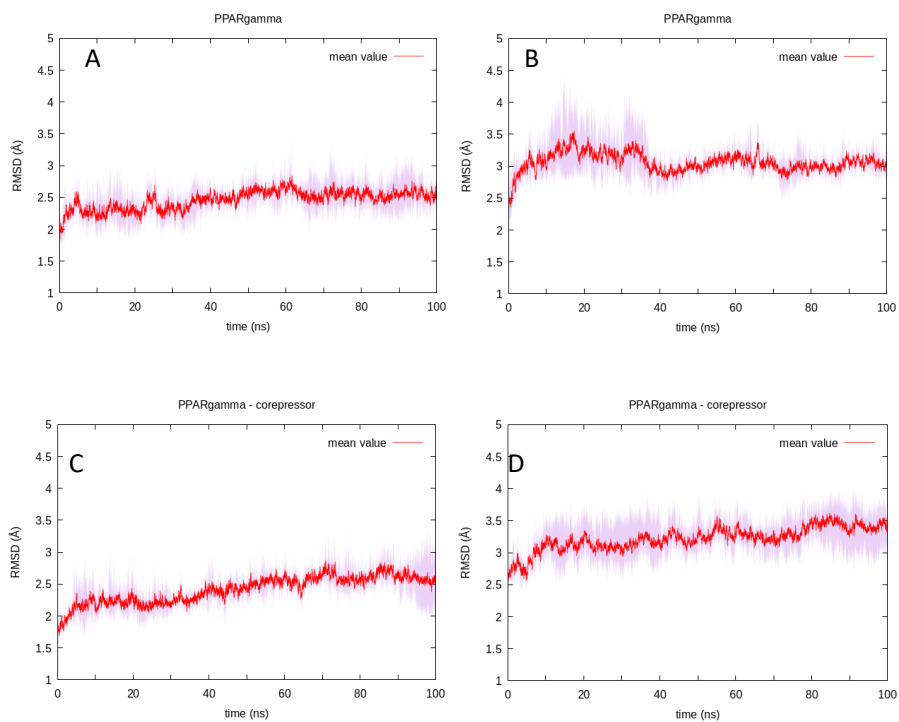

Figure S4. RMSD of PPARγ LBD by AA simulation (A), PPARγ LBD by Drude-2019 simulation (B). PPARγ LBD bound to the corepressor peptide by AA simulation (C); PPARγ LBD bound to the corepressor peptide by Drude-2019 simulation (D). The mean value of three replicas is represented as a red line and the variability is represented by the shaded region.

**Salt-bridge interatomic distance time series**

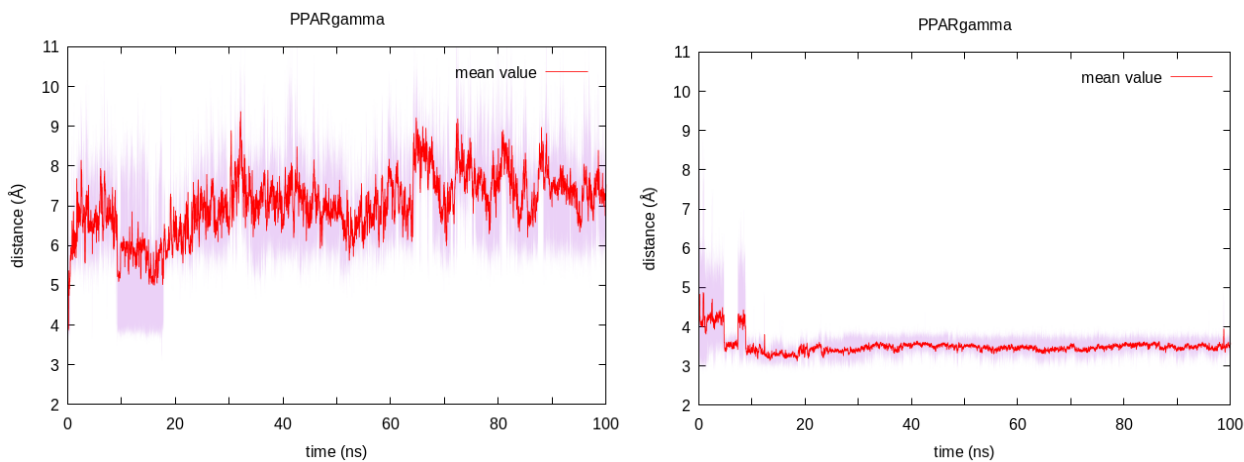

Figure S5A. Timeseries for the salt-bridge interatomic distance of the PPARγ LBD apo form, in the AA (left) and Drude-2019 simulations (right). The distance is calculated between two heavy atoms: CG of D411 residue, and NE2 of H453 residue.

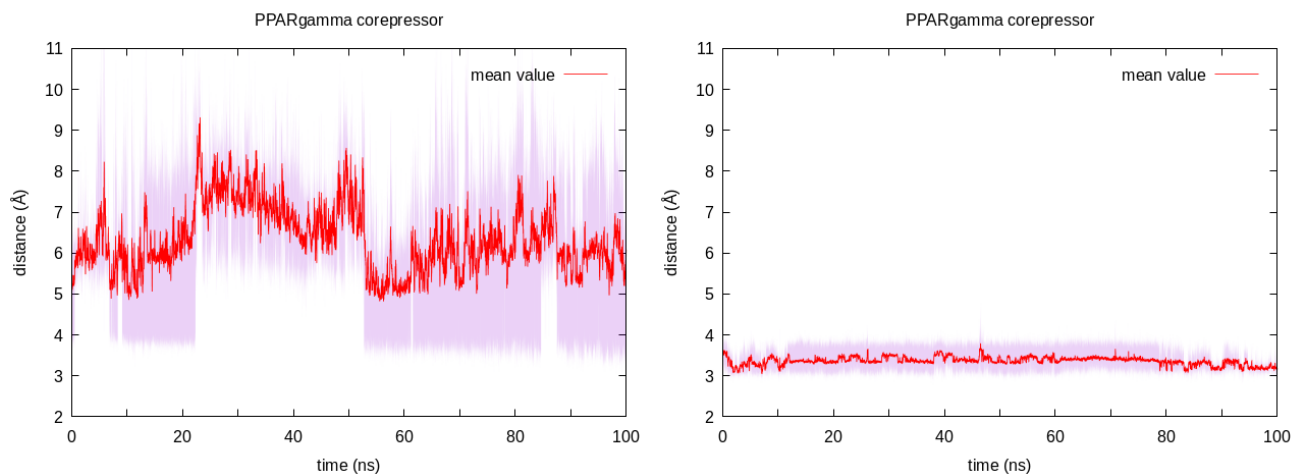

Figure S5B. Timeseries for the salt-bridge interatomic distance of the PPARγ LBD corepressor-bound form, in the AA (left) and Drude-2019 simulations (right). The distance is calculated between two heavy atoms: CG of residue D411, and NE2 of residue H453.

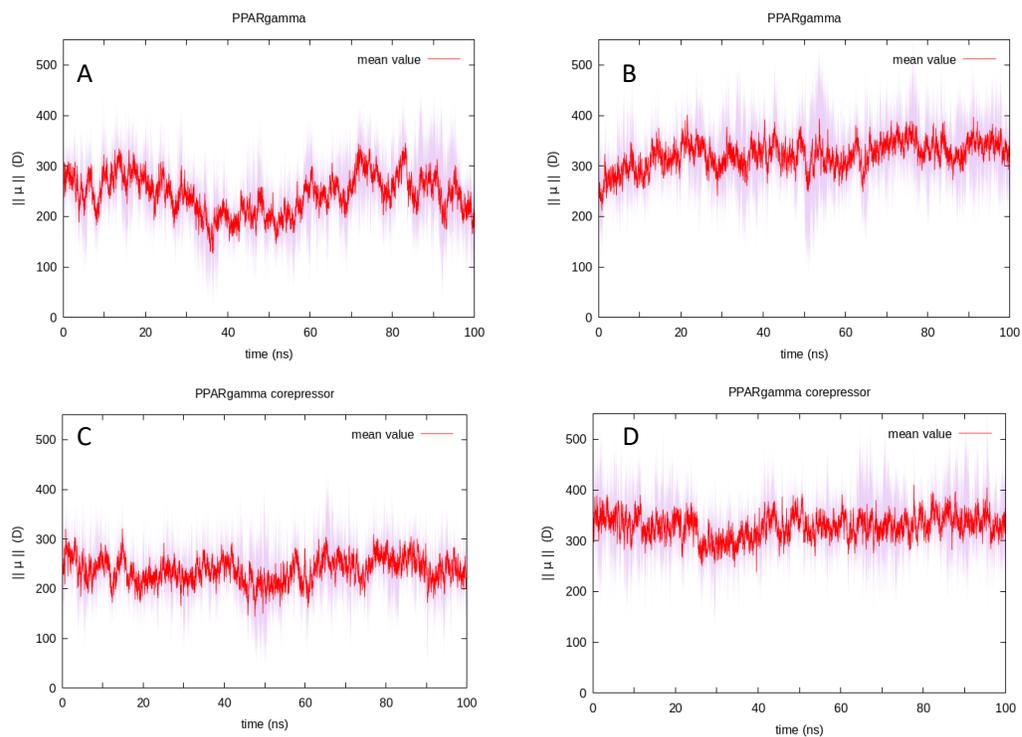

Figure S6. Protein dipole moment timeseries of the PPARγ LBD by AA simulation (A), PPARγ LBD by Drude-2019 simulation (B). PPARγ LBD bound to the corepressor peptide by AA simulation (C); PPARγ LBD bound to the corepressor peptide by Drude-2019 simulation (D). The mean value of 3 replicas is represented as a red line.

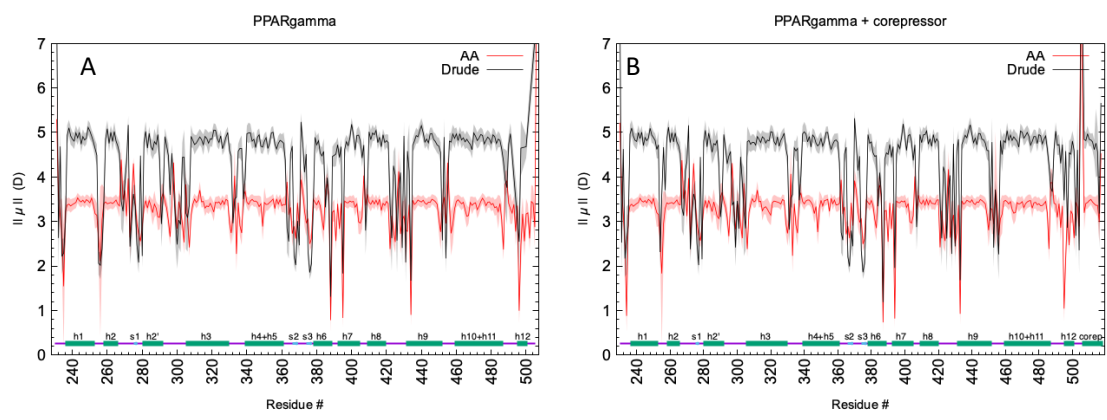

Figure S7. By-residue dipole moments of PPARγ for the apo protein (A) and for the LBD in complex with the corepressor peptide (B). Secondary structure elements are shown on the x axis: alpha helices (h1 – h12) as green, and beta strands (s1 – s3) as blue rectangles.

**Scatterplot of the correlation coefficients for PPARγ corepressor peptide complex**

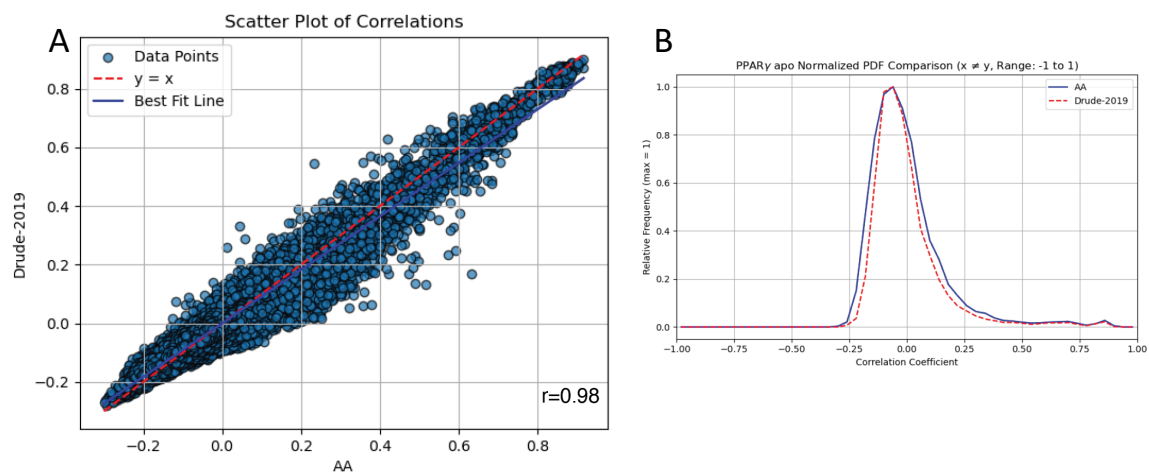

Figure S8. (A) Scatterplot of the correlation coefficients used to calculate the correlation maps from the molecular dynamics simulation of PPARγ corepressor peptide (Fig. 10B of the manuscript); from the AA simulations (x-axis) vs from the Drude-2019 simulations (y axis). The dashed line shown in red is x=y; the solid blue line is the best fit to the data excluding the self-correlation coefficients which equal 1, (B) the probability distribution of the correlation coefficients from the AA simulation (blue) and the Drude-2019 simulation (red) excluding the self-correlation coefficients which equal 1.

## Section 2: Community Network Analysis

Table S1. Composition of the nodes from the community network analysis of ubiquitin. Size represents the number of amino acids in a node. Node members are given by their amino acid numbering in the structure.

| Ubiquitin_AA | | | Ubiquitin_Drude-2019 | | |
|---|---|---|---|---|---|
| node id | size | members | node id | size | members |
| 1 | 9 | c(1 :4, 14 :18) | 1 | 7 | 1 :7 |
| 2 | 9 | 5 :13 | 2 | 11 | 8 :18 |
| 3 | 18 | c(19:24, 49:60) | 3 | 3 | 19 :21 |
| 4 | 12 | 25 :36 | 4 | 12 | 22 :33 |
| 5 | 14 | c(37:42, 69:76) | 5 | 8 | 34 :41 |
| 6 | 7 | c(43:48, 68) | 6 | 12 | c(42:44, 68:76) |
| 7 | 7 | 61 :67 | 7 | 8 | 45 :52 |

Table S2. Composition of the nodes from the community network analysis of PPARγ. Isoform PPARγ2 numbering of LBD residues: 230 - 505. Corepressor peptide is numbered from 506 - 517. Size represents the number of amino acids in a node. Node members are given by their amino acid numbering in the structure.

| PPAR_AA_apo | | | PPAR_AA_corep | | |
|---|---|---|---|---|---|
| node id | size | members | node id | size | members |
| 1 | 23 | c(230:234, 432:449) | 1 | 28 | c(230:232, 431:455) |
| 2 | 22 | 235:256 | 2 | 22 | 233:254 |
| 3 | 17 | 257:273 | 3 | 19 | 255:273 |
| 4 | 45 | c(274:296, 363:384) | 4 | 32 | c(274:289, 372:378) |
| 5 | 14 | c(297, 307:319) | 5 | 38 | c(290:308, 487:505) |
| 6 | 27 | c(298:306, 488:505) | 6 | 24 | 309:332 |
| 7 | 13 | 320:332 | 7 | 17 | 333:349 |
| 8 | 25 | c(333:348, 423:431) | 8 | 13 | 350:362 |
| 9 | 14 | 349:362 | 9 | 13 | 379:391 |
| 10 | 21 | 385:405 | 10 | 16 | 392:407 |
| 11 | 17 | 406:422 | 11 | 23 | 408:430 |
| 12 | 15 | 450:464 | 12 | 31 | 456:486 |
| 13 | 23 | 465:487 | 13 | 12 | 506:517 |

| PPAR_Drude-2019_apo | | | PPAR_Drude-2019_corep | | |
|---|---|---|---|---|---|
| node id | size | members | node id | size | members |
| 1 | 25 | 230:254 | 1 | 25 | 230:254 |
| 2 | 20 | 255:274 | 2 | 20 | 255:274 |
| 3 | 33 | c(275:284, 362:384) | 3 | 25 | c(275:277, 362:383) |
| 4 | 21 | 285:305 | 4 | 13 | 278:290 |
| 5 | 26 | 306:331 | 5 | 14 | 291:304 |
| 6 | 30 | 332:361 | 6 | 27 | 305:331 |
| 7 | 22 | 385:406 | 7 | 30 | 332:361 |
| 8 | 15 | 407:421 | 8 | 23 | 384:406 |
| 9 | 28 | 422:449 | 9 | 16 | 407:422 |

| 10 | 17 | 450:502 |  | 10 | 31 | 423:453 |
|----|----|---------|--|----|----|---------|
| 11 | 39 | 467:505 |  | 11 | 34 | 454:487 |
|    |    |         |  | 12 | 18 | 488:505 |

**Discussion of nodes in the PPARγ LBD.**

The specific compositions of the nodes in PPARγ LBD are given in the above tables, the discussion of the results for the PPARγ apo LBD are shown in Fig. 12 of the manuscript. The nodes correspond largely to entire secondary structure elements, mostly helices, however, the AA simulations have four nodes encompassing spatially adjacent residues belonging to different helices: the first community regroups the N-terminal residues with residues from H9 (node 1, 23 residues, 230:234, 432:449), the second node regroups H2', the ω loop, the beta sheet and helix 6 (node 4, 45 residues, 274:296, 363:384), the third one regroups the ω loop C-terminal residues with H12 residues (node 6, 27 residues, 298:306, 488:505), and the fourth one regroups the loop between H3 and H4, along with H4 and the H8 – H9 loop (node 8, 25 residues, 333:348, 423:431)). Node 4 is the one most coupled to other nodes in the apo structure. Interestingly, node 4 shows a relatively weak direct coupling to node 6, which contains the transcriptionally important H12, but it is strongly coupled to node 5, which encompasses the N-terminal end of H3. This lack of strong direct coupling may be due to the fact that the spatially near loop in node 6 is quite flexible. There is also a relatively strong coupling between the loop H8-H9 (node 8) with the rest of the protein. Interestingly, this loop is known to interact with cyclin H in other nuclear receptors, in particular RARα[12]. The PPARγ LBD is known to interact with cyclin D[13] in the context of regulating adipogenesis. The nodes encompassing the loop regions at either end of Helix 9 are well connected to the rest of the protein and are known to be important in the allostery related to phosphorylation in other receptors[12,14,15]. The small beta sheet appears as a node. Beta sheets are known from several studies to be involved in information transfer through correlated motions. In the AA simulations, we see that it is well connected to other important regions of the protein.

The CNA of the Drude-2019 simulation of the apo PPARγ LBD generally shows smaller nodes than those observed in the AA simulation. The largest node, node 3 (residues 275:284, 362:384) encompasses the residues of H6, the β sheet and some of the ω loop; the equivalent node in the AA simulation is node 4, however, the node from the Drude-2019 simulation is smaller. Many of the other nodes are along secondary structure elements. As in the AA simulation, there is no direct coupling between N3 and the helix 12 region of PPAR. In the Drude-2019 results, the coupling passes through 3 to 4 nodes depending on the path, while in the AA simulation, the coupling was either direct (weak) or through just one additional node. In addition to the couplings being different between the AA and the Drude-2019 simulation, the results show that the coupling between different regions of the PPAR ligand binding domain is less strong in the Drude-2019 simulations than in the AA simulations.

For the PPARγ corepressor complex, the CNA identified 13 community nodes for correlated motions from both the AA and the Drude-2019 simulations (Fig. S6). In both apo and corepressor bound cases, the protein has 12 nodes while the corepressor peptide forms its own node. More of the nodes identified in the AA simulations include sequentially distant, but spatially near resides (nodes 1, 4 and 5), while in the Drude-2019 simulation, there is only one node that includes sequentially distant, but spatially near residues (node 3). In the AA system, there are three nodes which connect to neighboring nodes: the first groups N-terminal residues with H9 (node 1, 230:232, 431:455), the second groups the N-terminal of the ω loop with the β sheet (node 4, 32 residues, 274:289, 372:378), and the third one associates the ω loop with H12 (node 5, 38 residues 290:308, 487:505). In the Drude-2019 simulation, node 3 encompasses the β sheet and the terminal end of the ω loop.

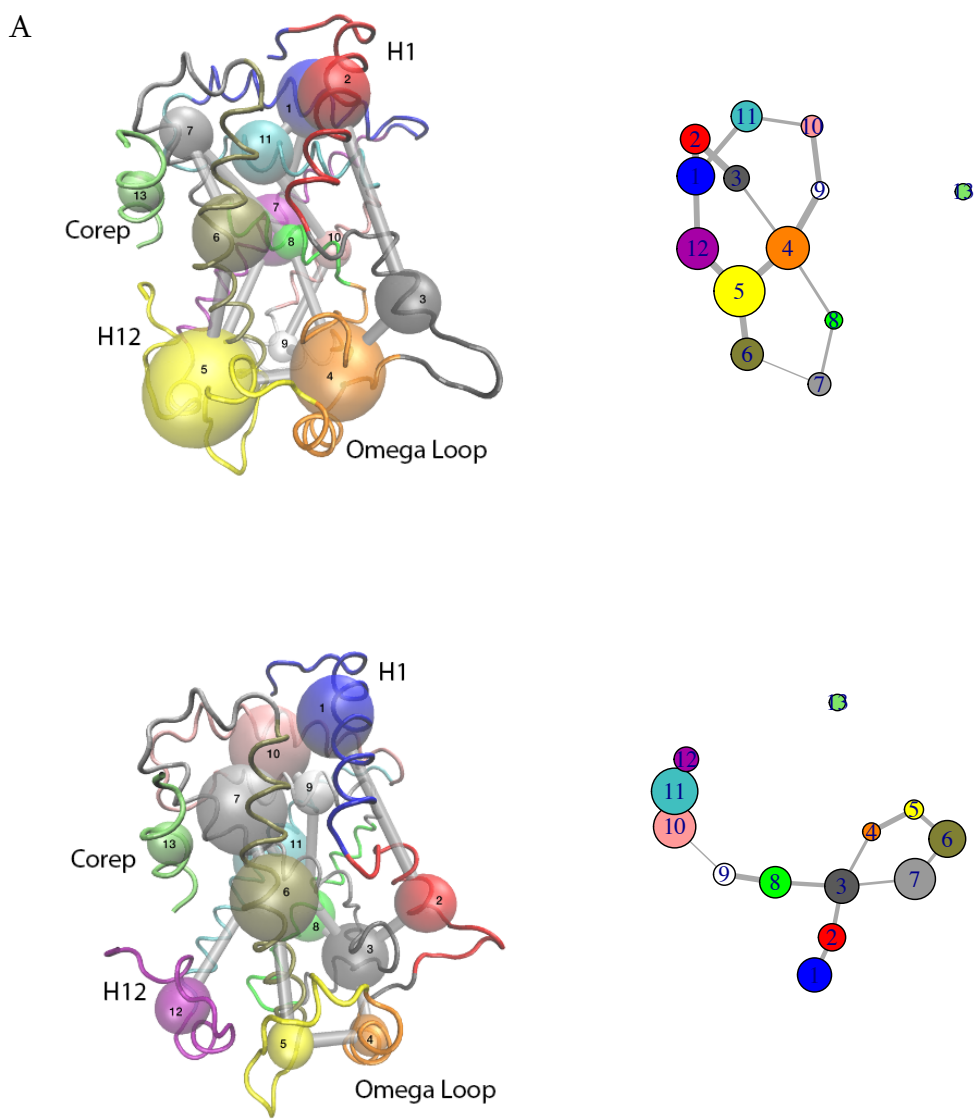

Figure S9. Community network analysis of PPARγ in complex with corepressor peptide. On the left, the colored nodes are superposed on the protein backbone structure, represented as a tube and colored according to the nodes. The edges are denoted as grey connections between the nodes, where the thickness indicates the strength of the correlation between two nodes. On the right is the network representation.   Representation on the PPARγ LBD corepressor bound form, from AA simulation (A), and Drude-2019 simulation (B).

The CNA analysis for both the AA and the Drude-2019 simulations show the corepressor peptide as a single node. And, we further notice that this node does not form any edges to any nodes in the ligand binding domain of PPARγ. Looking at the correlation plots of AA simulations, we see positive correlation between the corepressor peptide residues and the N-terminal of H4, while the Drude-2019 simulations did not capture these correlations. The correlations were weak and, as they did do not go over the 0.5 threshold, they are not represented by an edge. The addition of the corepressor peptide in the AA and Drude-2019 simulations does not seem to disrupt the community of the H3 – 4 loop and H4 residues, which forms the corepressor binding platform. In the AA results, the presence of the peptide seems to increase the H12 and ω loop community, passing from 27 (node 6) to 38 (node 5) residues, and reinforcing their correlations. In the Drude-2019 simulations, the addition of the peptide seems to decouple two different communities. First, the H11 – H12 community is split into two separate ones, connected by an edge (from node 11 to nodes 11 and 12). The second community, built around the ω loop (node 4), is divided into 2 separate nodes (nodes 4 and 5), connected by edges. Other nodes do not seem to be affected by the corepressor addition.

One significant distinction between the simulations using the two different force fields concerns the community that represents helix H12. In the AA simulations of both apo and corepressor-bound systems, H12 and part of the ω loop are coupled and are therefore represented by one community. This node is of medium size, with 27 residues for the apo form and with 38 residues for the corepressor-bound form. Furthermore, in both systems simulated with AA FF, the N-terminal residues of PPARγ are grouped in the same community with H9 residues, while in the Drude-2019 simulations, these N-terminal residues are in the same community as H1 residues. This coincides with high RMSF values for the N-terminal residues of the LBD in both Drude-2019 simulated systems, Fig. 9 of the manuscript. The lack of high correlating communities and the presence of communities largely representative of individual alpha-helices is apparent in the correlation maps, where the Drude-2019 simulations display attenuated correlations.

As there is overall a good correlation between the correlation coefficients obtained by the AA and Drude-2019 FF (see Figures 10 of the main text and Figure S8 above), we investigated whether lowering the threshold cutoff on the CNA analysis specifically in the Drude simulations would bring the CNA analysis of the Drude correlated motions more in line with the results from the AA. We focused this analysis on the apo PPARγ. We question, in particular, the difference between the AA and Drude simulation results for the coupling between the PPARγ omega loop and Helix H12. The default cutoff value is 0.5, here we tried 0.45 and 0.40. In both cases, no tendency towards convergence of the AA results (shown in Fig. 11 of the main text) and the Drude simulation results at different cutoffs is observed, Fig. S10. Even at lower cutoff values, no direct communication between H12 and the omega loop is observed. This indicates that simple modifications of the CNA parameters do not suffice to bring both force field to yield identical analysis of the correlation data.

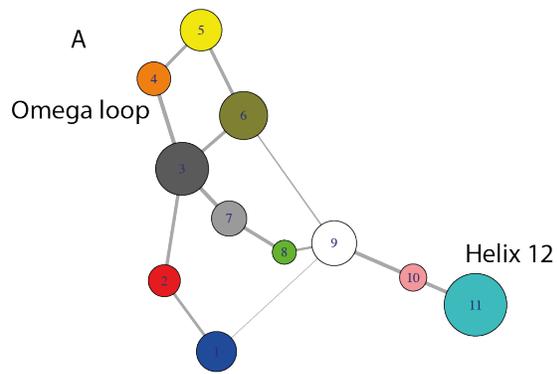
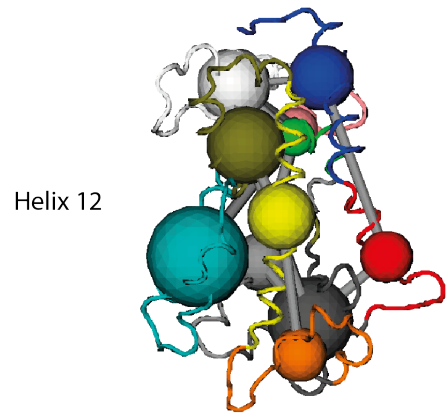

Threshold cutoff = 0.5 (default value)

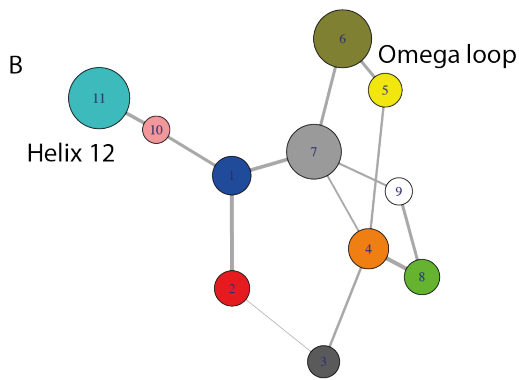
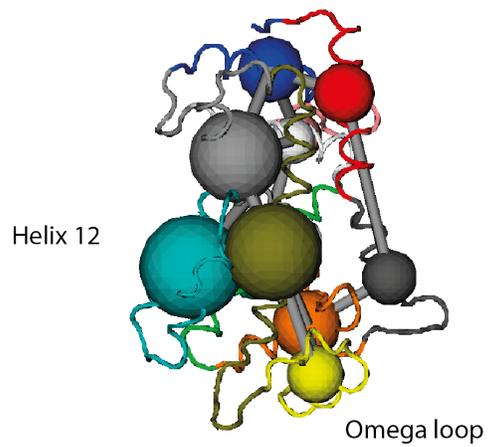

Threshold cutoff = 0.45

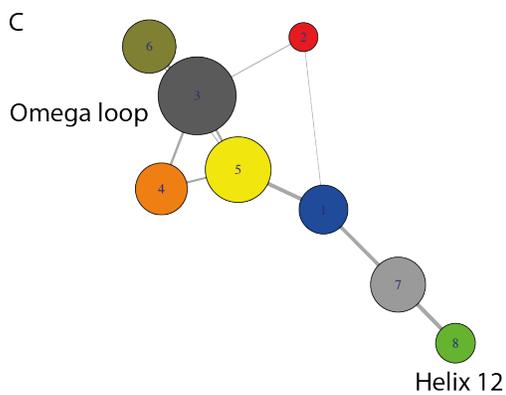
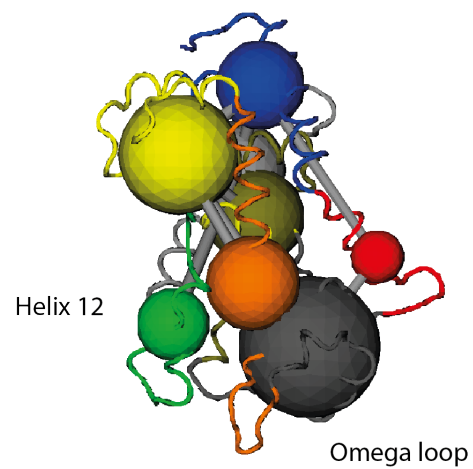

Threshold cutoff = 0.40

Figure S10: CNA analysis results for apo-PPARγ. Comparison of results using different threshold values. (A) cutoff value of 0.5, the default value used in this study, (B) cutoff value of 0.45, (C) cutoff value of 0.4.

**Section 3: Shortest Path Method (SPM)**

The Shortest Path Method (SPM) was used to assess the importance of individual residues and their pairwise connections in the structural dynamics of the two proteins[10]. This is in contrast to the community network analysis, which establishes communities around multiple residues. The SPM method produces a network graph based on mean distances and correlation values, and computes shortest path lengths using Dijkstra algorithm[16]. The shortest path is the most direct path with the most significant connection between two residues and shows how the residues are connected by the structural dynamics of the protein. The tool is mostly aimed at exploring key residues implicated in enzymatic activity, but here we use as a way to assess the similarities and differences of simulations using different force fields. The online tool provided at https://spmosuna.com was used with default values.

Concerning ubiquitin, the SPM analysis using the correlated motions calculated from the AA and Drude-2019 simulations show a similar pattern (Fig S7). Interestingly, in both cases, there is a path detected between the beginning of loop β1-β2 and the loop after the C-terminus of the α-helix. These regions of ubiquitin have been identified in other works as exhibiting a pincer-type motion. For this small, tightly packed protein, there do not seem to be any noticeably difference between the AA and Drude-2019 simulations in terms of the SMPs.

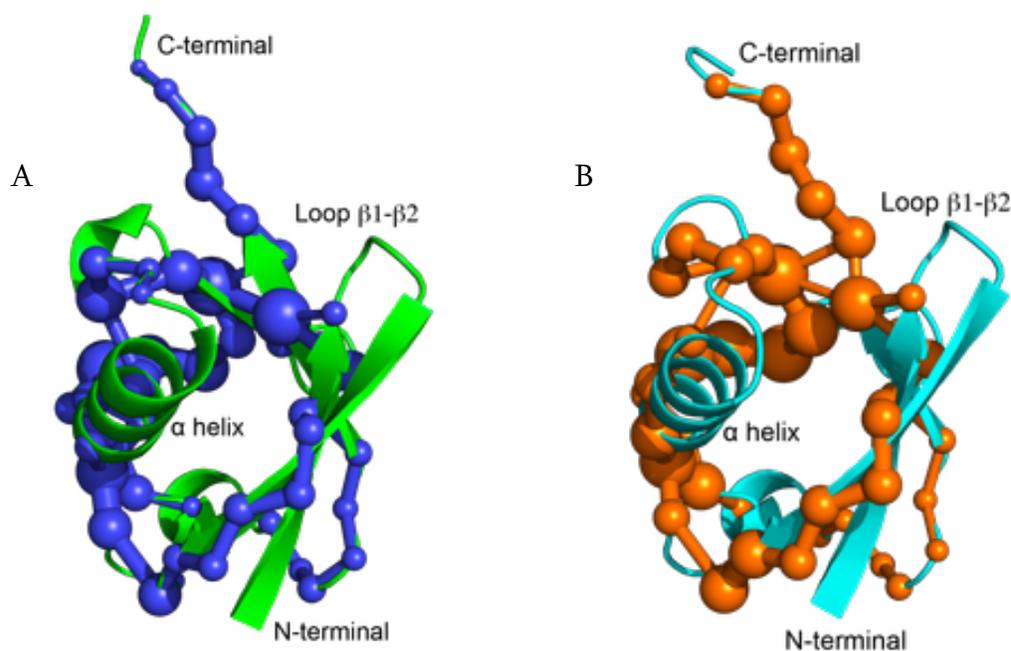

Fig S11. Shortest Path Method ball and stick representation mapped on ubiquitin average structure. The AA simulation result is in blue (A) and the Drude-2019 simulation result is in orange (B).

The PPARγ apo system (Fig. S8) shows a graph network connecting different nodes corresponding to same secondary structure elements. For example, if we look at the side view of the structure from AA simulation, we can see a path spanning the entire helix H10-11, and then continuing connecting the loop and H12, and even further, the ω loop. This suggests a correlation and coupling of these secondary structure elements. On the other hand, the Drude-2019 simulation shows no such connection and the functionally important H12 is not coupled to ω loop movements. Similar observations were made from the community network analysis.

Compared to the apo PPARγ, the SPM paths of PPARγ bound to the corepressor peptide are relatively different for both the AA and Drude-2019 calculations (Fig. S9). In this case, we again discern in the case of the AA simulation, the SPM spanning throughout the 'upper' region of the LBD and the one in the 'bottom' region with respect to the illustration, where H12 and the ω loop are connected. Interestingly, we see short paths between alpha helices.

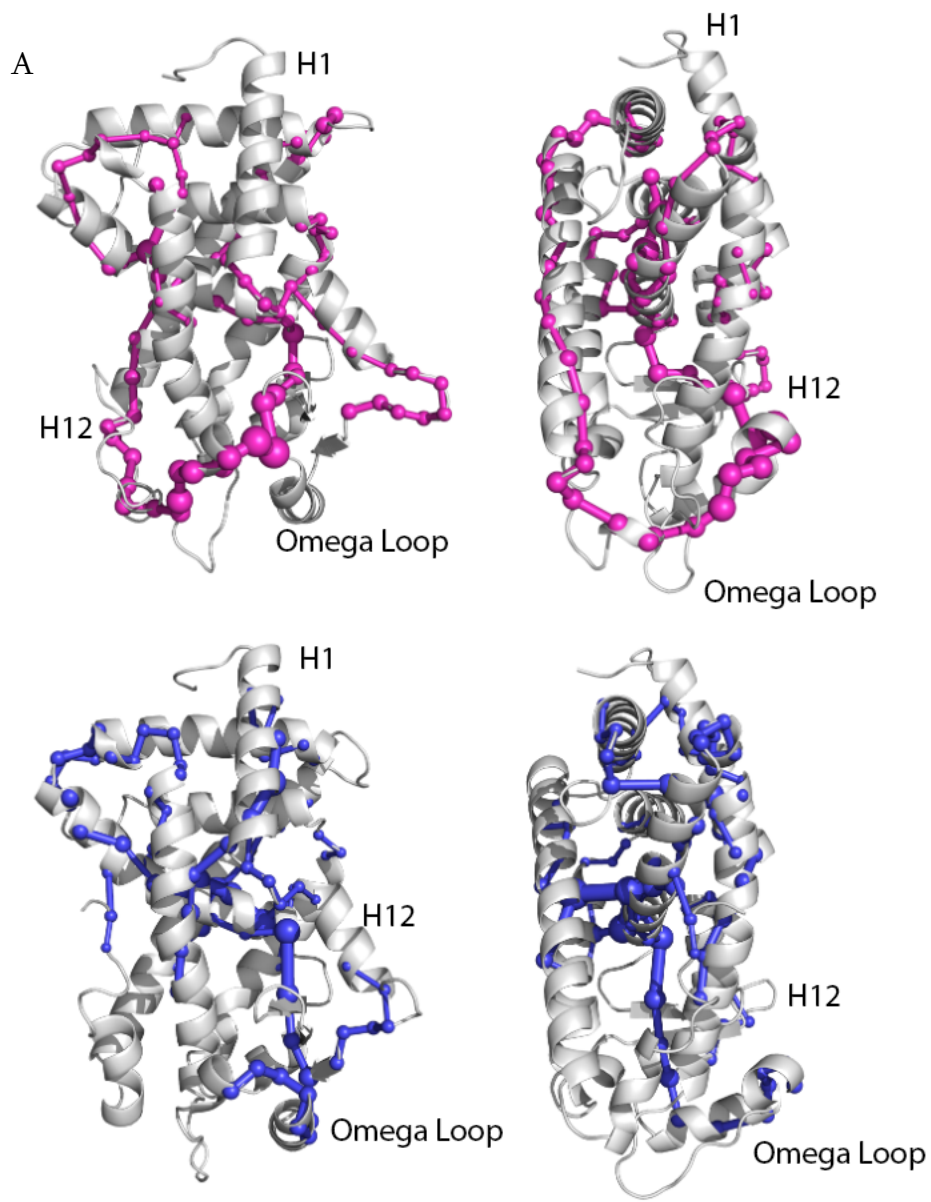

Figure S12. Shortest Path Method ball and stick representation mapped on the front and side views of PPARγ LBD apo form. The AA simulation result is in magenta (A) and the Drude-2019 simulation result in blue (B).

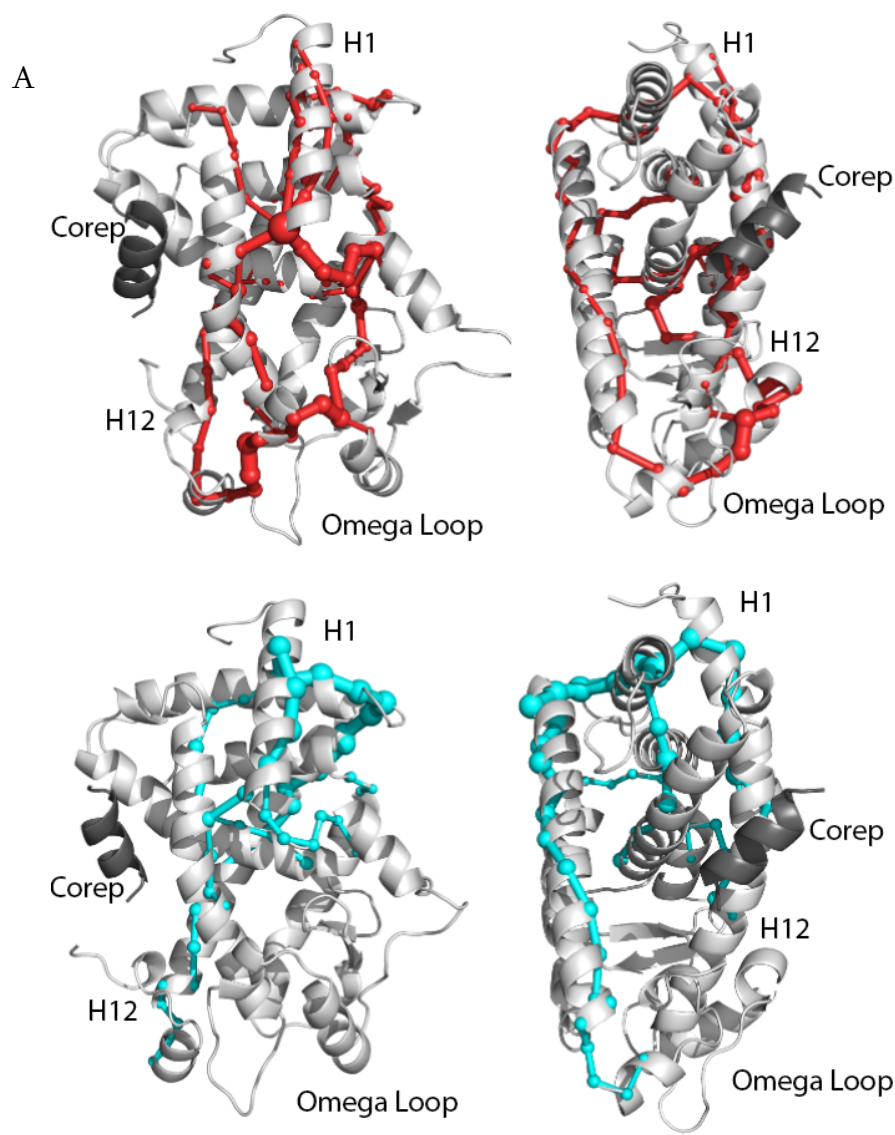

Figure S13. Shortest Path Method ball and stick representation mapped on the front and side views of PPARγ LBD with corepressor peptide bound. The AA simulation result is in red (A) and the Drude-2019 simulation result in cyan (B).

The use of the Drude-2019 force field leads to a decoupling of H12 and the ω loop region in the apo protein; the same observation was made from the community network analysis.

The corepressor peptide, even though it was included in the SPM network calculation, does not appear to participate in the shortest path representation. Despite the correlated motions between the corepressor peptide and helices H3 and H4 (Fig. 9B in the main manuscript), the co-repressor peptide is not connected to the rest of the protein in this analysis. A similar conclusion was made from the community network analysis, that is, the corepressor peptide does not enter into any communication network. We also notice the absence of the SPM path in the regions of the loop H3 – H4, probably caused by the addition of the corepressor peptide. This suggests that the presence of the peptide, while not directly implicated in a network, will perturb the underlying communication network of PPARγ.

**Section 4: Hydrophobic ILV cluster analysis**

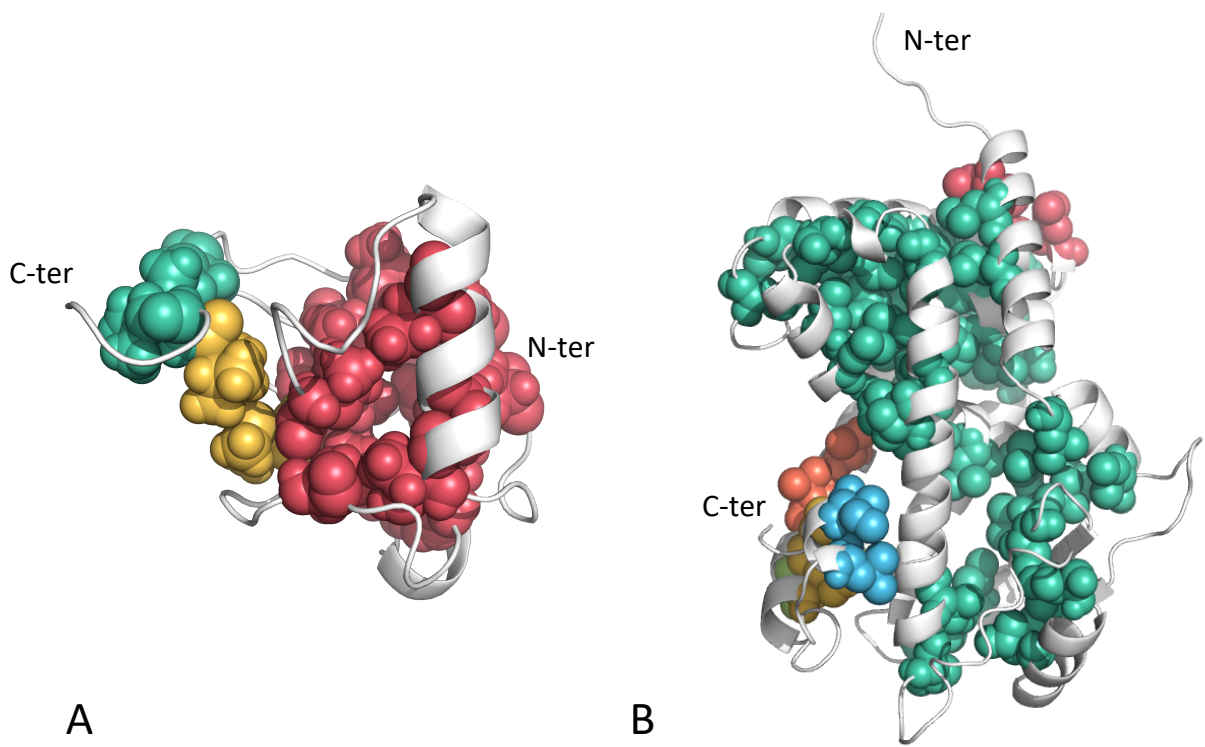

Figure S14. Hydrophobic clusters determined using the ProteinTools webserver using the default settings. A) Ubiquitin, the cluster 0 (red), cluster 1 (green) and cluster 2 (yellow). Cluster 0 was used for the energetic analysis of the molecular dynamics trajectories, B) PPARγ, cluster 0 (red), cluster 1 (green), cluster 2 (yellow), cluster 3 (orange), cluster 4 (not visible), cluster 5 (blue). Cluster 1 was used for the energetic analysis of the molecular dynamics trajectories.

**CHARMM commands used to define the clusters**

**Ubiquitin**

!cluster 0
define cluster0 sele ((resid 3:3 .or. resid 5:5 .or. resid 13:13 .or. resid 15:15 .or. resid 17:17 .or. resid 23:23 .or. resid 26:26 .or. resid 30:30 .or. resid 43:43 .or. resid 50:50 .or. resid 56:56 .or. resid 61:61 .or. resid 67:67) .and.
(.not. (type H* .or. type D* .or. type LP* ))) end

!cluster 1
define cluster1 sele ((resid 8:8 .or. resid 71:71) .and. -
(.not. (type H* .or. type D* .or. type LP* ))) end

!cluster 2
define cluster2 sele ((resid 44:44 .or. resid 70:70) .and. -
(.not. (type H* .or. type D* .or. type LP* ))) end

**PPARγ (apo and with corepressor peptide)**

!cluster 0
define cluster0 sele (resi 211:211 .or. resid 419:419 .or. resid 423 :423) .and. (.not. (type H* .or. type D* .or. type LP* )) end

!cluster 1
define cluster1 sele (resi 214:214 .or. resid 218:218 .or. resid 228:228 .or. resid 236:236 .or. -
 resid 237:237 .or. resid 249:249 .or. resid 255:255 .or. resid 262:262 .or. resid 277:277 .or. -
 resid 281:281 .or. resid 293:293 .or. resid 296:296 .or. resid 303:303 .or. resid 309:309 .or. -
 resid 317:317 .or. resid 318:318 .or. resid 322:322 .or. resid 325:325 .or. resid 326:326 .or. -
 resid 330:330 .or. resid 333:333 .or. resid 339:339 .or. resid 340:340 .or. resid 341:341 .or. -
 resid 353:353 .or. resid 356:356 .or. resid 377:377 .or. resid 379:379 .or. resid 384:384 .or. -
 resid 386:386 .or. resid 388:388 .or. resid 390:390 .or. resid 391:391 .or. resid 392:392 .or. -
 resid 393:393 .or. resid 400:400 .or. resid 403:403 .or. resid 406:406 .or. resid 409:409 .or. -
 resid 413:413 .or. resid 414:414 .or. resid 417:417 .or. resid 421:421 .or. resid 431:431 .or. -
 resid 435:435 .or. resid 436:436 .or. resid 442:442) .and. (.not. (type H* .or. type D* .or. type LP* )) end

!cluster 2
define cluster2 sele (resid 453:453 .or. resid 456 :456) .and. (.not. (type H* .or. type D* .or. type LP* )) end

!cluster 3
define cluster3 sele (resid 446:446 .or. resid 450 :450) .and. (.not. (type H* .or. type D* .or. type LP* )) end

!cluster 4
define cluster4 sele (resid 452:452 .or. resid 455 :455) .and. (.not. (type H* .or. type D* .or. type LP* )) end define cluster5 sele (resid 469:469 .or. resid 472 :472) .and. (.not. (type H* .or. type D* .or. type LP* )) end

**Section 5: Statistical analysis of hydrophobic native contacts results**

**Ubiquitin**

Wilcoxon rank-sum test statistic: 367694524.0
p-value: 0.0
The two datasets are statistically different (p < 0.05).

**AA**
Mean: 0.93
Median: 0.92
Standard Deviation: 0.06
Variance: 0.00
Skewness: -0.68
Kurtosis: 0.18
Shapiro-Wilk Test: Statistic=0.81, p-value=0.0
The data does not follow a normal distribution.

**Drude-2019**
Mean: 0.79
Median: 0.81
Standard Deviation: 0.08
Variance: 0.00
Skewness: -0.03
Kurtosis: -0.25
Shapiro-Wilk Test: Statistic=0.95, p-value=0.0
The data does not follow a normal distribution.

**PPARγ_apo**
Wilcoxon rank-sum test statistic: 1412048402915.0
p-value: 0.0
The two datasets are statistically different (p < 0.05).

**AA**
Statistics for 0.75:
Mean: 0.86
Median: 0.88
Standard Deviation: 0.07
Variance: 0.00
Skewness: -0.17760292273505357
Kurtosis: -0.10
Shapiro-Wilk Test: Statistic=0.93, p-value=0.0
The data does not follow a normal distribution.

**Drude-2019**
Statistics for 0.57:
Mean: 0.65
Median: 0.67
Standard Deviation: 0.08
Variance: 0.01
Skewness: 0.02
Kurtosis: -0.20
Shapiro-Wilk Test: Statistic=0.95, p-value=0.0
The data does not follow a normal distribution.

**PPARγ_corepressor**
Wilcoxon rank-sum test statistic: 1009764539147.0
p-value: 0.0
The two datasets are statistically different (p < 0.05)

**AA**
Statistics for 0.81:
Mean: 0.87
Median: 0.88
Standard Deviation: 0.06
Variance: 0.00
Skewness: -0.23
Kurtosis: 0.04
Shapiro-Wilk Test: Statistic=0.92, p-value=0.0
The data does not follow a normal distribution.

**Drude-2019**
Statistics for 0.67:
Mean: 0.84
Median: 0.87
Standard Deviation: 0.08
Variance: 0.01
Skewness: -0.15
Kurtosis: -0.17
Shapiro-Wilk Test: Statistic=0.94, p-value=0.0
The data does not follow a normal distribution.

**Section 6: Cluster Energy analysis**

We calculated the van de Waals and electrostatic energies of the main clusters in each system, the results are presented as box plots in Fig. S15. We see that, for ubiquitin, the van der Waals self-energy of the cluster in the Drude-2019 simulation is less negative than in the AA simulation, suggesting that an increased van der Waals energy reflects a disruption of the hydrophobic cluster (Fig. S15A, upper panel). Similar observations are seen in the cluster self-energy from the apo-PPARγ and PPARγ corepressor bound form simulations (Fig. S15B,C upper panels). Again, the van der Waals self-energies of the LIV clusters are significantly less negative in the Drude-2019 simulations than in the AA simulations. The differences are statistically significant as the Wilcoxon test gives a p value of zero. When adding the electrostatics contribution to the van der Waals, for the same clusters, we see that there is some compensation in the total self-energy, as the energy of the Ubiquitin cluster is now more negative using the Drude. This is however not the case for the PPARγ simulations (Fig. S15 ABC, lower panels). The balance of van der Waals and electrostatics thus differs between the two force fields, but it is not possible to relate it in a simple manner to the loss of native contacts observed in the simulations (Figure 13 of the main text). A full understanding of the impact of these force field changes on the correlation and allosteric coupling analysis would require further investigations.

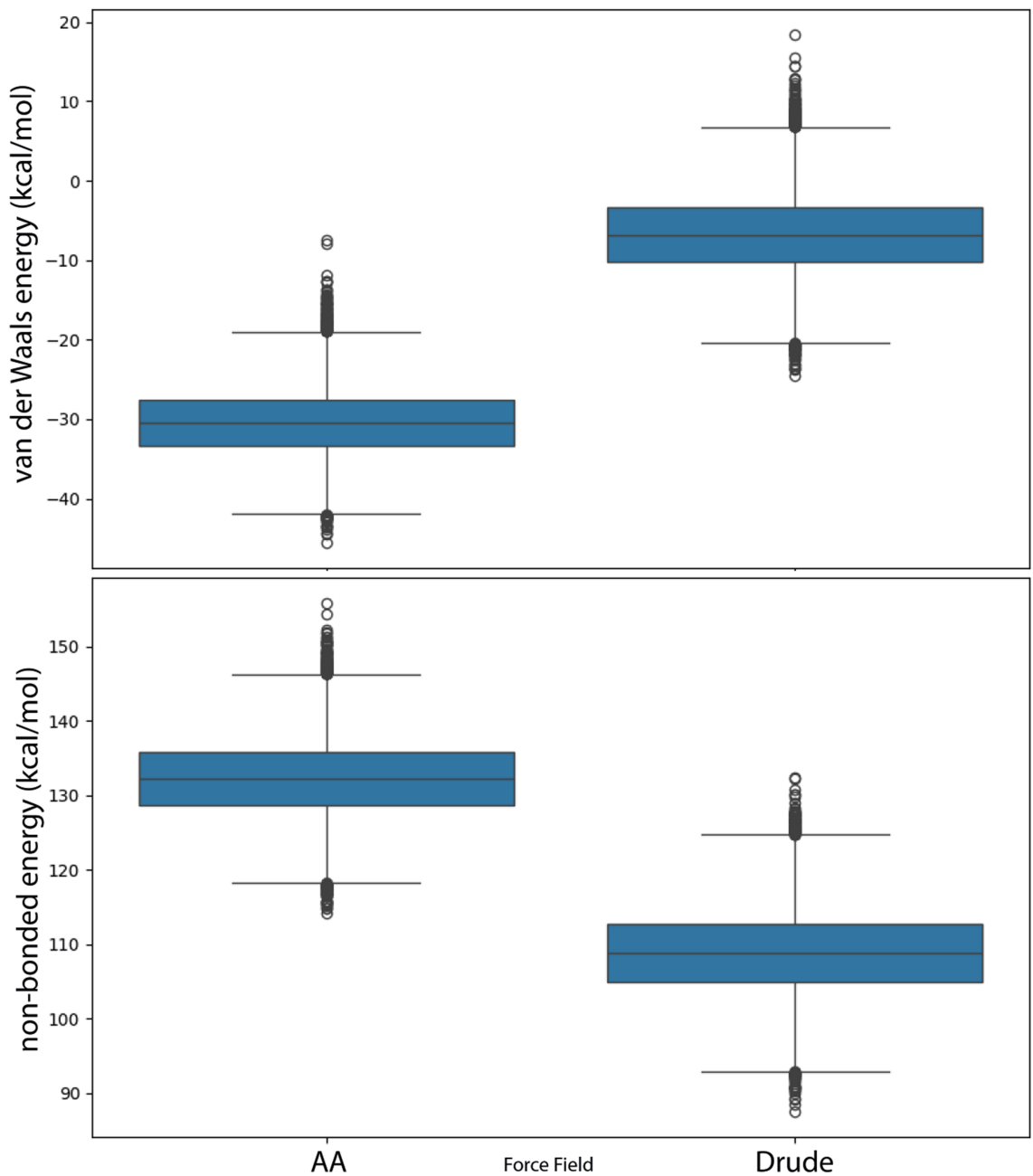

Figure S15. Ubiquitin. Top Figure, the original plot of the van der Waals energies the of hydrophobic cluster from the AA and Drude simulation; bottom Figure, the box plot of the van der Waals plus electrostatic energies of the hydrophobic cluster.

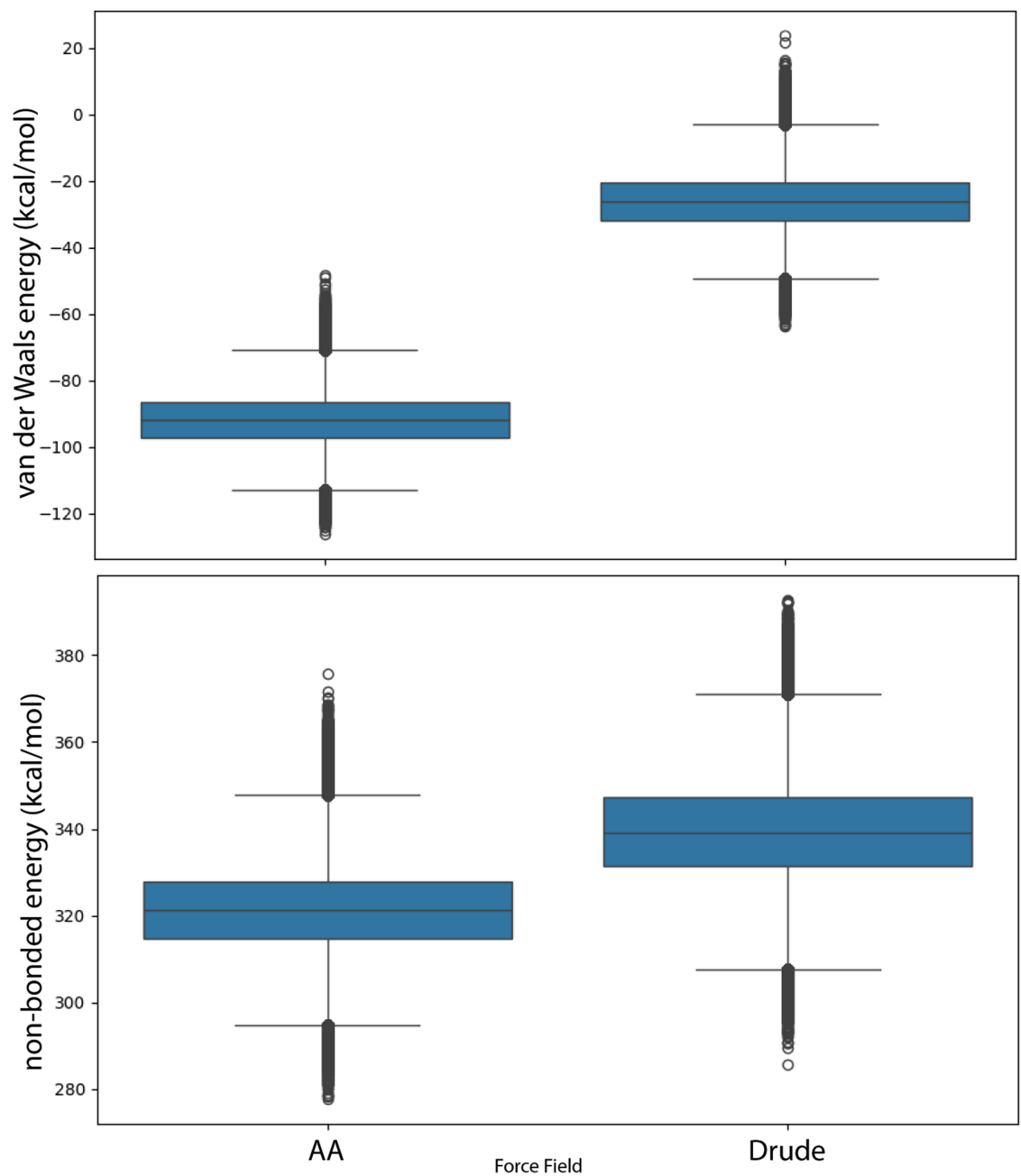

Figure S16 apo-PPAR. Top Figure, plot of the van der Waals energies the of hydrophobic cluster from the AA and Drude simulation; bottom Figure, the box plot of the van der Waals plus electrostatic energies of the hydrophobic cluster.

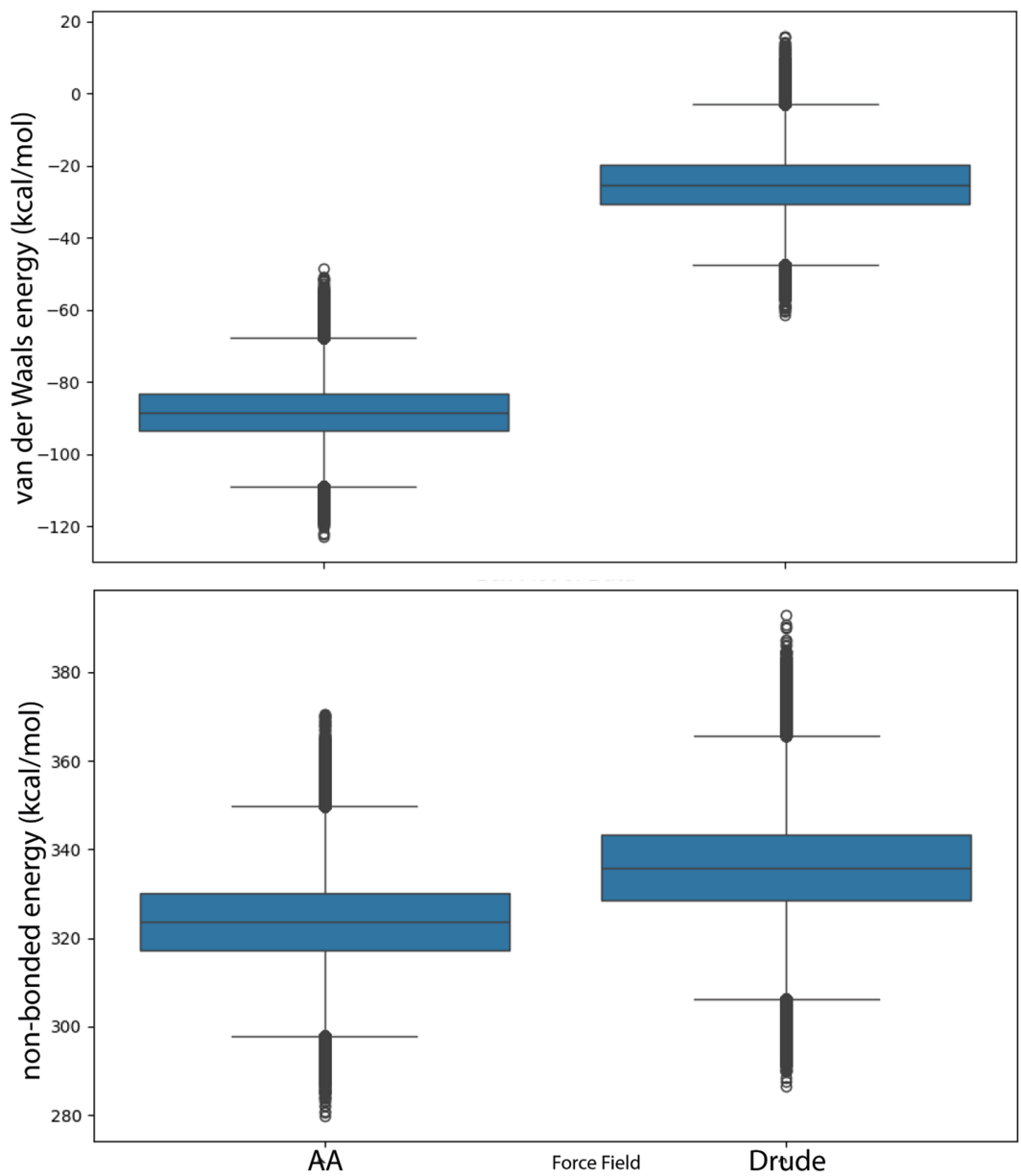

Fig S17. corep-PPAR. Top Figure, plot of the van der Waals energies the of hydrophobic cluster from the AA and Drude simulation; bottom Figure, the box plot of the van der Waals plus electrostatic energies of the hydrophobic cluster.

## Statistical analysis of hydrophobic cluster self-energy (van der Waals energies) results

### Ubiquitin

Wilcoxon rank-sum test statistic: 96132.0
p-value: 0.0
The two datasets are statistically different (p < 0.05).

**AA**
Statistics for -36.18:
Mean: -30.36
Median: -30.51
Standard Deviation: 4.30
Variance: 18.45
Skewness: 0.22
Kurtosis: 0.12
Shapiro-Wilk Test: Statistic=0.10, p-value=8.93e-19
The data does not follow a normal distribution.

**Drude-2019**
Statistics for -2.22:
Mean: -6.73
Median: -6.84
Standard Deviation: 5.09
Variance: 25.86
Skewness: 0.18
Kurtosis: 0.10
Shapiro-Wilk Test: Statistic=0.10, p-value=8.08e-15
The data does not follow a normal distribution.

### PPARγ_apo

Wilcoxon rank-sum test statistic: 18160.0
p-value: 0.0
The two datasets are statistically different (p < 0.05).

**AA**
Mean: -91.84
Median: -91.99
Standard Deviation: 7.84
Variance: 61.46
Skewness: 0.11
Kurtosis: 0.02

Shapiro-Wilk Test: Statistic=0.10, p-value=3.45e-21
The data does not follow a normal distribution.

**Drude-2019**
Statistics for -31.27:
Mean: -26.04
Median: -26.18
Standard Deviation: 8.63
Variance: 74.53
Skewness: 0.09
Kurtosis: 0.01
Shapiro-Wilk Test: Statistic=1.00, p-value=1.0
The data seems to follow a normal distribution.

**PPARγ_corepressor**

**AA**
Wilcoxon rank-sum test statistic: 12449.0
p-value: 0.0
The two datasets are statistically different (p < 0.05).
Statistics for -77.63:
Mean: -88.27
Median: -88.40
Standard Deviation: 7.65
Variance: 58.48
Skewness: 0.10
Kurtosis: 0.025
Shapiro-Wilk Test: Statistic=0.10, p-value=5.05e-08
The data does not follow a normal distribution.

**Drude-2019**
Statistics for -20.35:
Mean: -25.16
Median: -25.30
Standard Deviation: 8.27
Variance: 68.45
Skewness: 0.10
Kurtosis: 0.02
Shapiro-Wilk Test: Statistic=0.10, p-value=1.08e-13
The data does not follow a normal distribution